\documentclass[reprint,onecolumn,superscriptaddress,showpacs,preprintnumbers,nofootinbib,aps,prd]{revtex4-1}
\usepackage{graphicx,color}
\usepackage{graphicx}
\usepackage{amsmath,amssymb,amsfonts}
\usepackage{stackrel}
\usepackage{enumitem}
\usepackage[english]{babel}
\usepackage{verbatim}
\usepackage{comment}
\usepackage{ulem}
\usepackage{xcolor}
\usepackage{subcaption}
\usepackage{dcolumn}
\usepackage{physics}
\usepackage{verbatim}
\usepackage{float}
\usepackage{color}
\usepackage{multirow}
\usepackage{graphicx}
\usepackage{epstopdf}
\usepackage{epsfig}
\usepackage{slashed}
\usepackage{physics}
\usepackage{multirow}
\usepackage{makecell}
\usepackage{placeins}
\usepackage{setspace}
\usepackage{tabularx}
\usepackage{colortbl}
\usepackage{float}
\usepackage{placeins}

\definecolor{darkred}{rgb}{.7,.1,.1}
\usepackage{hyperref}
\hypersetup{
    pdfnewwindow=true,  
    colorlinks=true,     
    linkcolor=darkred,     
    citecolor=blue,      
    filecolor=blue,      
    urlcolor=blue        
}
 
\graphicspath{
  {./}
  {figures/}
}

\definecolor{dark-green}{rgb}{0.1,0.7,0.3}

\newcommand{\tf}{\texorpdfstring}
\newcommand{\gev}{~\text{GeV}}
\newcommand{\tev}{~\text{TeV}}

\newcommand{\fbi}{~\text{fb}^{-1}}
\newcommand{\abi}{~\text{ab}^{-1}}

\def\nn{\nonumber}

\begin{document}
\title{Lepton flavor of four-fermion operator and fermion portal dark matter}

\author{Yuxuan He}
\email{heyx25@pku.edu.cn}
\affiliation{School of Physics and State Key Laboratory of Nuclear Physics and Technology, Peking University, Beijing 100871, China}

\author{Gang~Li}
\thanks{Corresponding author: \href{mailto:ligang65@mail.sysu.edu.cn}{ligang65@mail.sysu.edu.cn}}
\email{}
\affiliation{School of Physics and Astronomy, Sun Yat-sen University, Zhuhai 519082, P.R. China.}

\author{Jia Liu}
\thanks{Corresponding author: \href{mailto:jialiu@pku.edu.cn}{jialiu@pku.edu.cn}}
\email{}
\affiliation{School of Physics and State Key Laboratory of Nuclear Physics and Technology, Peking University, Beijing 100871, China}
\affiliation{Center for High Energy Physics, Peking University, Beijing 100871, China}

\author{Xiao-Ping Wang}
\thanks{Corresponding author: \href{mailto:hcwangxiaoping@buaa.edu.cn}{hcwangxiaoping@buaa.edu.cn}}
\email{}
\affiliation{School of Physics, Beihang University, Beijing 100083, China}
\affiliation{Beijing Key Laboratory of Advanced Nuclear Materials and Physics, Beihang University, Beijing 100191, China}

\author{Xiang Zhao}
\email{zhaox88@mail2.sysu.edu.cn}
\affiliation{School of Physics and Astronomy, Sun Yat-sen University, Zhuhai 519082, P.R. China.}

\begin{abstract}
\vspace*{0.5cm}

We study the ultraviolet realization of semileptonic four-fermion operator $O_{ledq}^{\alpha \beta 11}$ that incorporates Majorana dark matter (DM) in both lepton-flavor-conserving (LFC) and lepton-flavor-violating (LFV) scenarios at the one-loop level via box diagram, which effectively alleviates the lower bounds on the new physics scale. The interplay between the model-independent constraints on the Wilson coefficients and DM direct detection, relic density, and collider searches in the context of fermion portal DM model with two mediators is investigated. We find that both the projected future constraint on the LFC Wilson coefficient $C_{ledq}^{2211}/\Lambda^2< (12.3~\text{TeV})^{-2}$ from the measurements of neutrino non-standard interaction in the next-generation neutrino oscillation experiments, and LFV constraint $C_{ledq}^{1211}/\Lambda^2< \left(2.2\times 10^3~\text{TeV} \right)^{-2}$ from ongoing charged-lepton-flavor-violation searches, provide a complementary exploration of the parameter space encompassing the DM mass and scalar mass. With the colored mediator mass typically around $2~\text{TeV}$, the sensitivity of the indirect constraints on the four-fermion operator could surpass those of collider searches and DM direct detection, in scenarios where the masses of the DM and scalar are close. By ensuring the correct DM relic density, however, we obtain that the collider searches and DM direct detection are more sensitive to the electroweak scale DM and scalar compared to the indirect constraints.

\end{abstract}

\pacs{}
\maketitle

\section{Introduction} 
\label{sec:intro}

While the Standard Model (SM) of particle physics has achieved striking success, its lack of explanation of fundamental phenomena, such as the origins of neutrino masses and fermion flavor structure, and providing particle candidates for dark matter (DM) underscores the need for physics beyond the SM (BSM). In light of null results in searches for new resonances at the Large Hadron Collider (LHC), the effective field theory (EFT) framework offers a powerful way to explore new physics~\cite{Isidori:2023pyp}.

In the EFT approach, BSM interactions are systematically parameterized as a series of higher-dimensional operators with the corresponding Wilson coefficients suppressed with inverse powers of the new physics scale $\Lambda$.
At the lowest order with mass dimension 5, only one operator is present that gives masses to neutrinos~\cite{Weinberg:1979sa}, while exploding numbers of effective operators emerge at higher orders~\cite{Buchmuller:1985jz,Grzadkowski:2010es,Henning:2015alf}. 

There has been growing interest in dimension-6 
four-fermion operators, as being utilized to fit precision data and address anomalies observed at the LHC and in low-energy experiments~\cite{Falkowski:2017pss,Cirigliano:2023nol,Fernandez-Martinez:2024bxg,Karmakar:2024gla}, and intriguing connections with DM models~\cite{Cepedello:2023yao}.
Focusing on the semileptonic operators with two leptons and two quarks, the one-by-one constraints on the lepton-flavor-conserving (LFC) are rather stringent, and global fits are essential to mitigate or even eliminate the tensions in different observables~\cite{Cirigliano:2023nol}. 
On the other hand, for the lepton-flavor-changing (LFV) operators, possible tensions are milder since less experimental data is available~\cite{Fernandez-Martinez:2024bxg}.

Consider the dimension-6 four-fermion operator~\cite{Grzadkowski:2010es}
\begin{align}
\label{eq:dim6-operator}
    O_{ledq }^{\alpha\beta s t} &= ( \bar L_\alpha^j e_{R\beta} ) ( \bar d_{Rs}  Q_t^j )\;,
\end{align}
where $L=(\nu_L,e_L)^T$ and $Q=(u_L,d_L)^T$ denote the left-handed $SU(2)_L$ doublets of leptons and quarks, respectively. $j$ is the isospin index, and $\alpha$, $\beta$, $s$, $t$ are flavor indices.  
This operator 
is forbidden under the $U(3)^5$ flavor symmetry, and highly suppressed within minimal flavor violation hypothesis~\cite{DAmbrosio:2002vsn,Isidori:2023pyp}.
We will consider the first-generation quarks, $s=t=1$, and the lepton flavor indices $\alpha,\beta =1,2$.

In general, the stringent constraints on the Wilson coefficients of four-fermion operators challenge direct searches for ultraviolet (UV) physics. 
However, one-loop UV realizations of the four-fermion operators with the participation of DM particles in the box diagram can notably alleviate the restrictions on the mass scale $\Lambda$.
In the class of DM models~\cite{Cepedello:2023yao}, the $Z_2$ symmetry that stabilizes the DM prohibits tree-level contributions to the Wilson coefficients of four-fermion operators. Consequently, upon integrating out heavy new particles, the Wilson coefficients are suppressed by a factor of $1/(16\pi^2)$.
Moreover, the Wilson coefficients are proportional to $f_{\rm NP}^4$, where $f_{\rm NP}$ denotes new physics couplings. This dependency has the potential to further lower the new physics scale for small values of $f_{\rm NP}$. Previous studies of connections between charged-lepton-flavor-violation (cLFV) observables and DM were conducted in Refs.~\cite{Herrero-Garcia:2018koq,Toma:2013zsa,Vicente:2014wga}.

In this work, we will investigate the complementarities of indirect constraints on the Wilson coefficients of the four-fermion operator $O_{ledq}^{\alpha\beta 11}$ in both LFC and LFV scenarios from the measurements of neutrino non-standard interaction (NSI) and cLFV searches for $\mu \to e$ conversion, and DM relic density, direct detection, and collider searches in the context of fermion portal DM model~\cite{Bai:2013iqa,DiFranzo:2013vra,An:2013xka,Bai:2014osa} with two mediators and Majorana fermionic DM. 
We highlight that
\begin{itemize}
    \item In the LFC scenario, the anticipated future sensitivity to the neutrino NSI in next-generation neutrino oscillation experiments could play a nontrivial role in investigating new physics, while ensuring that tensions among various observables are eliminated.
    \item In the LFV scenario, the scattering of Majorana DM with nuclei can appear at tree level, and $\mu \to e$ conversion arises solely from the contribution of the box diagram at the one-loop level, with no other cLFV processes occurring at this order. 
    \item In both scenarios, the model-independent constraints on the Wilson coefficients of the four-fermion operator offer a distinctive probe of the fermion portal DM model, particularly in the parameter space where the masses of the DM and scalar are either close or large, depending on the requirement of DM relic density. 
\end{itemize}

The remainder of the paper is organized as follows. In Sec.~\ref{sec:model}, the fermion portal DM model is studied in detail. In Sec.~\ref{sec:wilson}, the Wilson coefficients of the four-fermion operator $O_{ledq}^{\alpha\beta 11}$ are calculated. In subsequent sections, the sensitivities in DM relic density, direct detection, and collider searches are investigated. In Sec.~\ref{sec:result}, the combined results are discussed in benchmark scenarios of masses and coupling with the relic density of $\chi$ satisfying $\Omega_{\chi} h^2 \leq 0.1199$ or $\Omega_{\chi} h^2 = 0.1199$. 
We conclude in Sec.~\ref{sec:conclusion}. More details are provided in Appendix~\ref{app:g-2} and Appendix~\ref{app:anapole} about the anomalous magnetic moment for leptons and DM direct detection with one-loop exchange of photon, respectively.

\section{Fermion portal dark matter model}
\label{sec:model}

We consider a simplified model that induces the four-fermion operator $O_{ledq }^{\alpha\beta 11}$ , incorporating Majorana fermionic DM $\chi$, which remains a more viable DM candidate compared to Dirac fermionic DM. 
In addition to the DM particle $\chi$, we introduce a fermion doublet $F = (F^+, F^0)^T$, scalar $S$, and colored mediator $\phi_d$, all of which are $Z_2$ odd. The quantum numbers of these fields under $\rm SU(3)_C \times SU(2)_L \times U(1)_Y \times Z_2$ are presented in Table~\ref{tab:quantum-numbers}, and are documented in the systematic classification of DM models for four-fermion operators~\cite{Cepedello:2023yao}.

\begin{table}[!htb]
\begin{center}
\renewcommand{\arraystretch}{1.2} 
\begin{tabular}{c|c|c|c|c}
\hline 
\hline
new fields & $\rm SU(3)_C$ & $\rm SU(2)_L$ & $\rm U(1)_Y$ & $~~~~\rm Z_2~~~~$\\ \hline
$\chi$ & $\mathbf{1}$ & $\mathbf{1}$ & 0 & $-1$ \\ \hline
 $F$ & $\mathbf{1}$ & $\mathbf{2}$ & $\frac{1}{2}$ & $-1$\\ \hline
 $S$ & $\mathbf{1}$ & $\mathbf{1}$ & $1$ & $-1$ \\ \hline
 $\phi_d$ & $\mathbf{3}$ & $\mathbf{1}$ & $-\frac{1}{3}$  & $-1$\\ \hline
\hline 
\end{tabular}
\end{center}
\caption{Quantum numbers associated with the new fields introduced in the fermion portal DM model.}
\label{tab:quantum-numbers}
\end{table}

The Lagrangian for this simplified model generating the four-fermion operator is given by
\begin{align}
\label{eq:UV-model}
\mathcal{L}&=f_{LS}(\bar{L} F_R) S^* + f_{\chi S}(\bar{\chi}_L e_{R}) S \nonumber\\
&+ f_{FQ}(\bar{F}_R Q) \phi_d^* + f_{d\chi}(\bar{d}_R \chi_L) \phi_d + \text{h.c.},
\end{align}
where $f_{LS}$, $f_{\chi S}$, $f_{FQ}$, and $f_{d\chi}$ are coupling constants.
The flavor indices are omitted, and the couplings are assumed to be real and positive for simplicity. 

In the presence of the dark mediators $S$ and $\phi_d$ in this fermion portal DM model~\cite{Bai:2013iqa,An:2013xka,DiFranzo:2013vra,Bai:2014osa}, the DM candidate $\chi$ interacts with SM particles, leading to distinct observables in direct detection, indirect detection, and collider experiments. In the subsequent sections, we will analyze each type of signature separately. 

\section{Wilson coefficients }
\label{sec:wilson}

\begin{figure}[!htb]
\centering
         \includegraphics[width=0.3\linewidth]{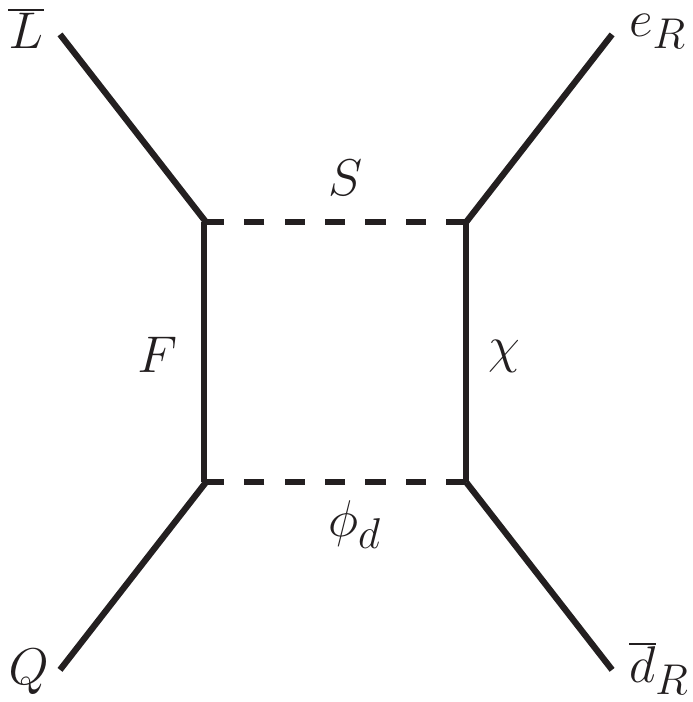}
\caption{{\it Dark loop}: box diagram with dark particles in the loop.  }
\label{fig:NSI-Maj}
\end{figure}

Given the interactions in Eq.~\eqref{eq:UV-model}, the one-loop diagram in Fig.~\ref{fig:NSI-Maj} can be generated. We refer to this type of box diagram as ``{\it dark loop}'', which was proposed to explain the flavor anomalies in $B$ physics~\cite{Huang:2020ris,Capucha:2022kwo,Cepedello:2022xgb}.

The Wilson coefficient of the effective operator $O_{ledq}^{\alpha \beta 11}$ is calculated using \texttt{Package-X}~\cite{Patel:2015tea, Patel:2016fam}, which is expressed as follows
\begin{align}
\label{wilson1}
& \frac{C_{ledq}^{\alpha \beta 11}}{\Lambda^2}=- \frac{f^4_{\rm NP}}{8 \pi^2}\times \left[ I(m_F^2,m_{\phi}^2,m_{S}^2,m_{\chi}^2) \right.\nn\\
&+\left.I(m_{S}^2,m_{\phi}^2,m_{F}^2,m_{\chi}^2) 
+  I(m_{\chi}^2,m_\phi^2,m_{F}^2,m_{S}^2)\right]\;,
\end{align}
where the effective coupling $f_{\rm NP}$ and the loop function $I(x,y,z,w)$ are defined as
\begin{align}
\label{eq:fNP-coupling}
    f_{\rm NP} &= \left(f_{LS} f_{\chi S} f_{FQ} f_{d\chi}\right)^{1/4}\;, \\
I(x,y,z,w) &= \dfrac{x^2 \log(x/y)}{4(x-y)(x-z)(x-w)}\;.
\end{align}
This result agrees with Ref.~\cite{Bischer:2018zbd} by taking $m_\chi = m_F = 0$.

\begin{figure}[t]
\centering
\centering
\begin{minipage}[t]{0.48\linewidth}
    \centering
    \includegraphics[width=0.9\textwidth]{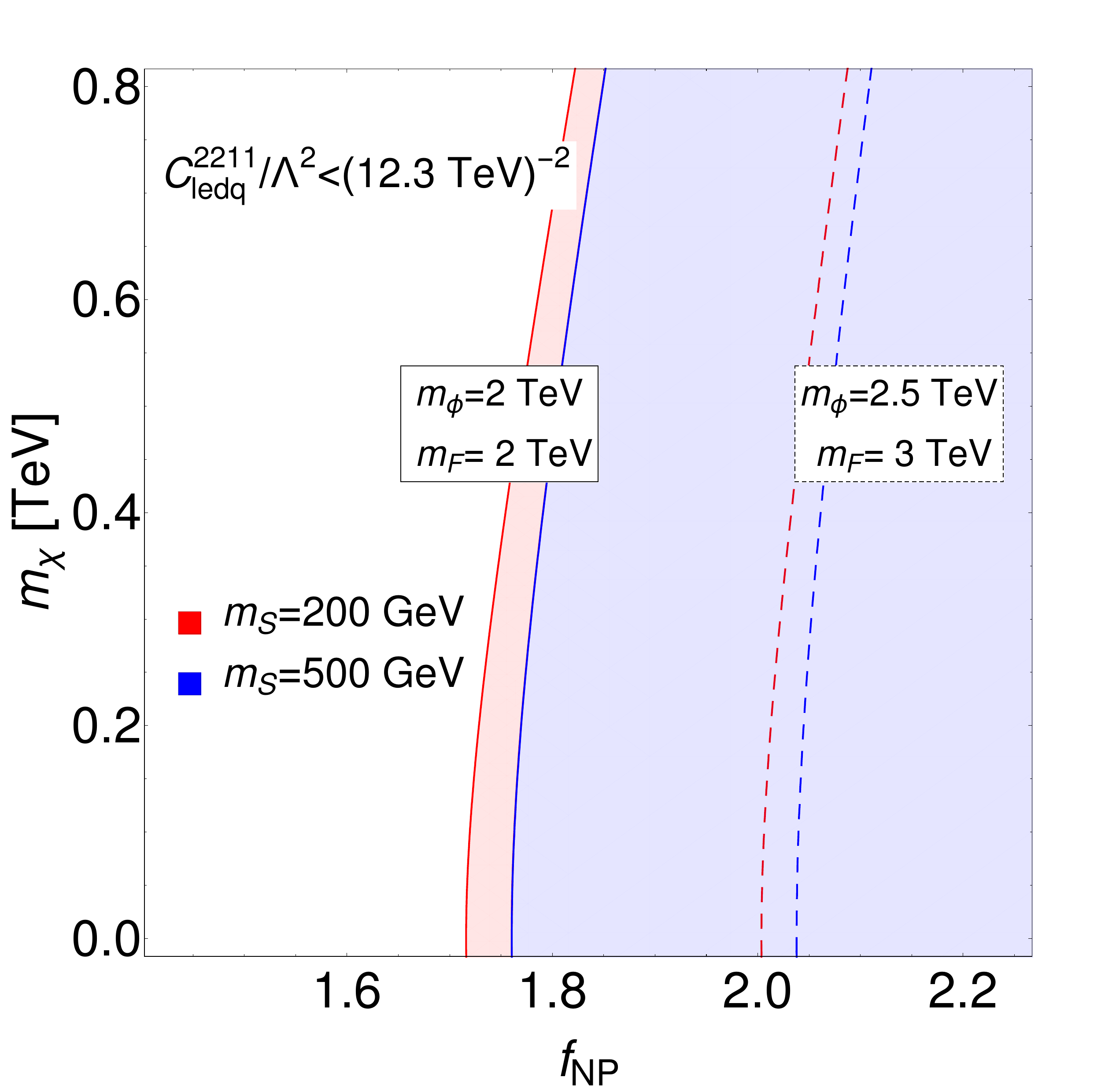}
\end{minipage}%
\begin{minipage}[t]{0.48\linewidth}
    \centering
    \includegraphics[width=0.9\textwidth]{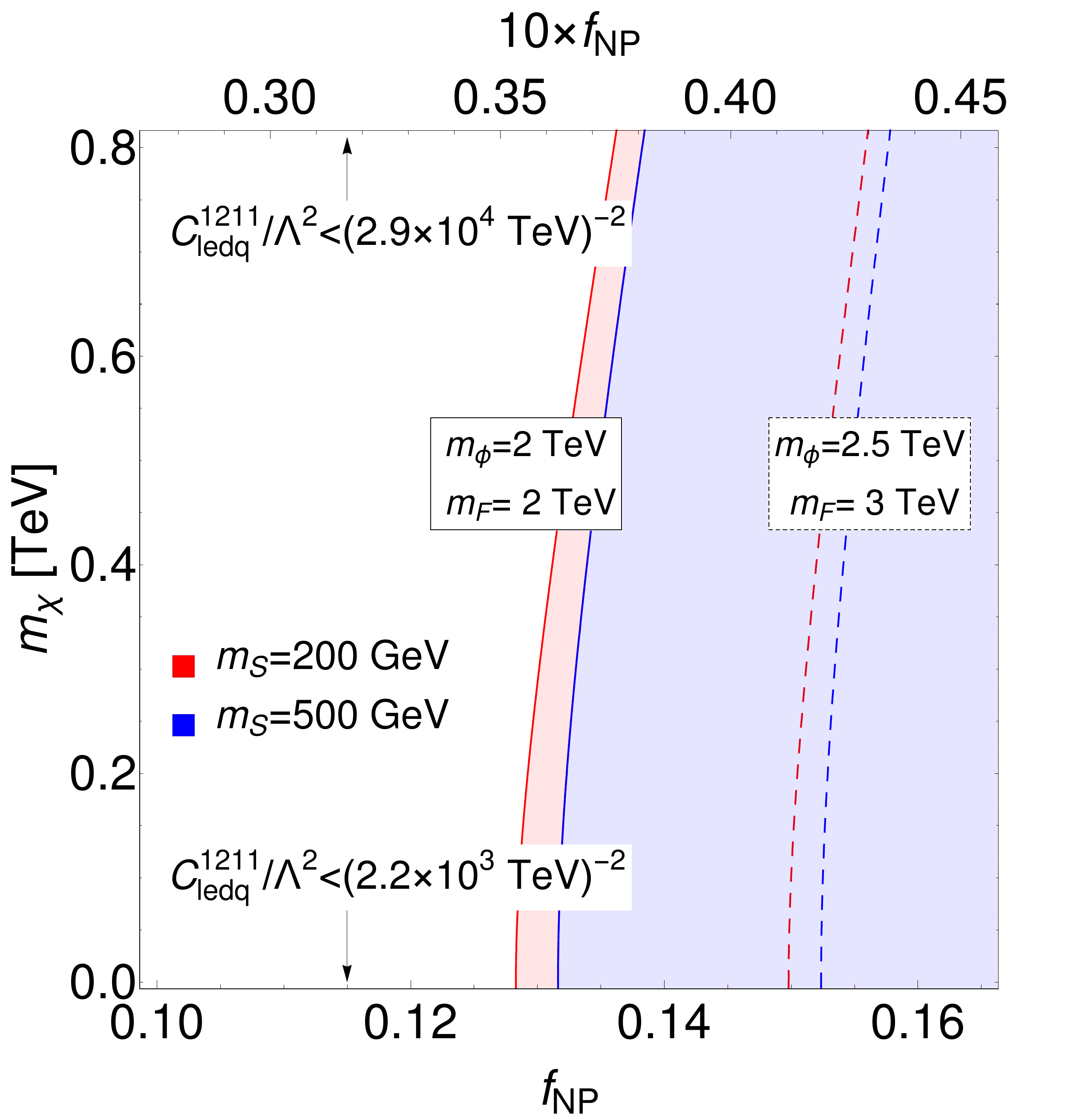}
\end{minipage}
     \caption{
     Contours of the Wilson coefficients as a function of $f_{\rm NP}$ and $m_\chi$ for the assumptions of the masses $m_\phi =2.5\tev~(2\tev)$,  and $m_F = 3\tev~(2\tev)$ 
      with solid (dashed) curves.
      The shaded regions in red (blue) color indicate exclusion for $m_S = 200\gev~ (500\gev)$.
     {\bf Left}: $ C_{ledq}^{22 11}/\Lambda^2< (12.3~\text{TeV})^{-2}$ is depicted for $f_{\rm NP}$ in the lower axis.
     {\bf Right}: $C_{ledq}^{12 11}/\Lambda^2< (2.2 \times 10^3$ $\text{TeV})^{-2}$  and $(2.9 \times 10^4~\text{TeV})^{-2}$ are depicted for $f_{\rm NP}$ in the lower and upper axes, respectively.
     }
     \label{Fig:wilson_contour} 
\end{figure}

The Wilson coefficient is constrained by searches in low-energy experiments and at colliders depending on the lepton flavor of the four-fermion operator. According to the global analysis in Ref~\cite{Cirigliano:2023nol}, the allowed value of the Wilson coefficient for the LFC operator $O_{ledq}^{2211}$ is $C_{ledq}^{2211}/\Lambda^2=\left(0.017\pm 0.039\right)\tev^{-2}$, which is consistent with zero within $1\sigma$.
Therefore, independent probes of the LFC operator are essential to disentangle the origin of new physics.
Refs.~\cite{Du:2020dwr,Du:2021rdg} have investigated the prospects of next-generation neutrino oscillation experiments in the search for NSIs~\cite{Proceedings:2019qno}, which are described by
certain SMEFT operators at the electroweak scale. These studies indicate the highest sensitivity for the operator $O_{ledq}^{2211}$, $\mid C_{ledq}^{2211}\mid /\Lambda^2<\left(12.3~\text{TeV}\right)^{-2}$~\cite{Du:2021rdg}. 
This enhanced sensitivity is attributed to the significant increase in neutrino production from pion decay, particularly in the presence of charged-current (CC) neutrino NSIs induced by $O_{ledq}^{2211}$ at lower energies.
Note that this operator does not interfere with the CC neutrino interactions in the SM, so the experimental measurements are blind to the sign of the Wilson coefficient. Henceforth, we will consider the magnitude of the Wilson coefficient $ C_{ledq}^{2211} /\Lambda^2$ with the absolute value symbol being omitted.

We illustrate the constraints on $f_{\mathrm{NP}}$ and $m_\chi$ from future LFC searches in the left panel of Fig.~\ref{Fig:wilson_contour} with two benchmark values of the scalar mass $m_S = 200\gev$ (red region) and $500\gev$ (blue region) for $m_\phi = m_F = 2\tev$ (solid curve), or  $m_\phi = 2.5\tev$, $m_F=3\tev$ (dashed curves). We find that this constraint does not strongly depend on $m_\chi$, but it is highly sensitive to $f_{\mathrm{NP}}$, which is expected to be in the range of approximately 1.7 to 2.1. It is important to note that there are currently no constraints from existing LFC searches, as the results are consistent with the SM predictions.

In the case of the LFV operator, the current constraint on the Wilson coefficient $C_{ledq}^{1211}/\Lambda^2< \left(2.2\times 10^3 \tev \right)^{-2}$~\cite{Fernandez-Martinez:2024bxg} is derived from charged-lepton-flavor-violation (cLFV) searches for $\mu \to e$ conversion in nuclei~\cite{Badertscher:1981ay,SINDRUMII:1993gxf,SINDRUMII:1996fti,SINDRUMII:2006dvw}. The projected limit  $ C_{ledq}^{1211} /\Lambda^2< \left(2.9\times 10^4 \tev \right)^{-2}$~\cite{Haxton:2024lyc} in the upcoming Mu2e~\cite{Mu2e:2014fns,Bernstein:2019fyh} and COMET~\cite{COMET:2018auw,COMET:2018wbw} experiments, representing an improvement by approximately two orders of magnitude. The cLFV searches\,\footnote{In our simplified model, the cLFV $\mu\to e$ conversion can be generated at one-loop level, while the other cLFV processes like $\mu \to e \gamma$ first arise at two-loop level.} are also insensitive to the sign of the Wilson coefficient of the operator $O_{ledq}^{1211}$, and we omit the absolute value symbol for convenience.

In the right panel of Fig.~\ref{Fig:wilson_contour}, we present the contours of LFV limits as functions of $f_{\rm NP}$ and $m_\chi$, by choosing the benchmark values of $m_S$, $m_F$, and $m_\phi$ as the left panel of Fig.~\ref{Fig:wilson_contour}. Unlike the LFC scenario, constraints from both the current and future experimental efforts are depicted. The current limit is shown on the lower axis, while the anticipated future limit is displayed on the upper axis. The current limit for $f_{\rm NP}$ will be around 0.13 to 0.15, with expectations that the future limit could improve by an order of magnitude, potentially reaching as low as 0.04. It is important to note that despite the stringent constraint on the LFV interaction, the BSM particles could potentially exist at the TeV scale or even lower, owing to the suppression of loop factor $\sim 1/(16\pi^2)$ and the coupling dependence $\sim f_{\rm NP}^4$.

Besides the effective interaction expressed as the four-fermion operator $\mathcal{O}_{ledq}^{\alpha\beta 11}$, the dark particles can also contribute to the anomalous magnetic momentum of leptons. To address the discrepancies between the experimental measurements and the SM predictions, improved SM calculations or other new physics contributions are needed; see Appendix~\ref{app:g-2} for details.

\section{DM relic density}
\label{sec:relic_density}

\begin{figure}[H]
\captionsetup[subfigure]{justification=centering}
\centering
         \includegraphics[width=0.2\linewidth]{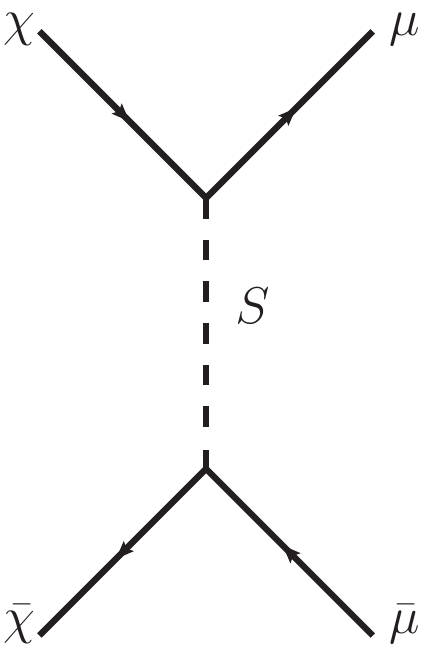}
\caption{The Feynman diagram for Majorana DM annihilation $\chi \bar \chi\to \mu^+ \mu^-$.
}
\label{fig:DMannhil-Maj}
\end{figure}

Given the Lagrangian in Eq.~\eqref{eq:UV-model}, the Majorana DM pair $\chi$ can annihilate to muon pair and $d$ quark pair, mediated by $S$ and $\phi_d$, respectively. In our setup, where \textbf{$m_{\phi} \gg m_S$} and the ratio $f_{d\chi}/f_{\chi S} \sim \mathcal{O}(1)$, the contribution from the process mediated by $\phi_d$, is negligible. 
Tree-level Feynman diagram for the annihilation $\chi \bar \chi\to \mu^+ \mu^-$ is depicted in Fig.~\ref{fig:DMannhil-Maj}. The corresponding thermal-averaged cross section is~\cite{Liu:2021mhn} 
\begin{align}
    \langle \sigma v \rangle
    = \frac{f_{\chi S}^4}{32 \pi} \frac{m_{\mu}^2}{m_S^4} \frac{1}{(1+x)^2}+v^2 \frac{f_{\chi S}^4}{48 \pi m_S^2} \frac{x\left(1+x^2\right)}{(1+x)^4}\;,
\end{align}
where $x \equiv m_\chi^2 / m_S^2$, and $v$ is the relative velocity of two DM particles, which is
typically around $0.3 c$ at the freeze-out temperature. 
Since $m_\mu^2/m_S^2 \ll v^2$, the $p$-wave contribution to the thermal-averaged annihilation cross section is dominant~\cite{Bai:2013iqa,Bai:2014osa,Liu:2021mhn}.

\begin{figure}[H]
\centering
\includegraphics[width=0.55\linewidth]{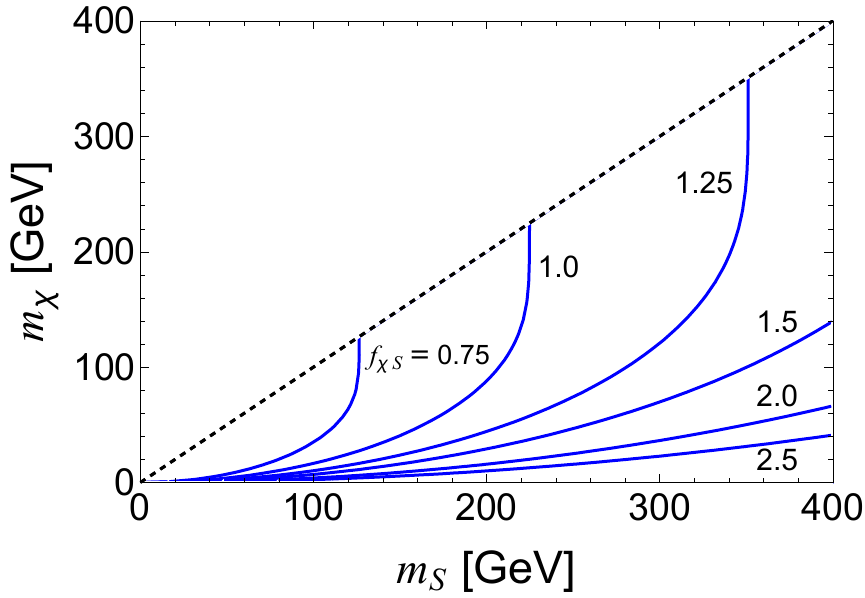}
\caption{The masses $m_S$ and $m_\chi$ for different choices of the coupling $f_{\chi S}$ after fitting the DM relic density. The contour plots are displayed only for $m_\chi< m_S$.
}
\label{fig:ann}	
\end{figure}

In Fig.~\ref{fig:ann}, we present the DM relic density contours for $m_S$ and $m_\chi$ with different choices of $f_{\chi S}$, considering the relic density $\Omega_{\chi} h^2 = 0.1199 \pm 0.0022$~\cite{Planck:2018vyg}.  Regions below the contour curves indicate overabundance. 

The DM annihilation channel $\chi \chi \to \mu^+ \mu^-$ is $p$-wave dominant and thus suppressed by the velocity of DM in the  whether during BBN or the recombination period. Current constraint on the cross section $\langle \sigma v \rangle$ for DM annihilation to muon pairs is $\langle \sigma v \rangle \leq 10^{-25}~\text{cm}^3/\text{s}$~\cite{Cirelli:2024ssz}
for DM mass around 100 GeV. This is a relatively loose limit, considering that the freeze-out annihilation cross section $\langle \sigma v \rangle_\text{freeze-out} \approx 10^{-26} ~\text{cm}^3/\text{s}$.  For light DM with mass below $\mathcal{O}(10)\gev$, indirect detection gives the constraint $\langle \sigma v \rangle \leq 10^{-26} ~\text{cm}^3/\text{s}$~\cite{Cirelli:2024ssz}. Similarly, the cross section is $p$-wave  suppressed by a factor of $ \sim 10^{-6}$, so that indirect detection can also not effectively constrain DM annihilation. Therefore, we can safely conclude that our model is free from indirect detection constraints.

\section{DM direct detection}
\label{DD}

\begin{figure}[H]
	\centering
\includegraphics[width=0.38\linewidth]{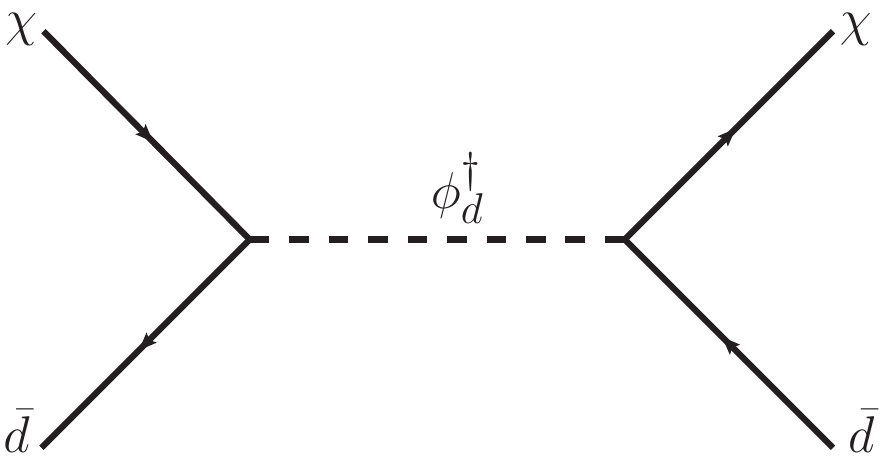}
	\caption{Feynman diagram for the tree-level scattering between DM and the SM quark.
  }
 \label{fig:MDD1}	
\end{figure}

The Majorana DM $\chi$ can interact with the SM quark through the $f_{d\chi}$ term in Eq.~\eqref{eq:UV-model}, giving rise to the tree-level scattering process depicted in Fig.~\ref{fig:MDD1}.

By using the Fierz identity~\cite{Bai:2013iqa} 
\begin{align}
 (\bar{\chi}_Ld_R)(\bar{d}_R\chi_L)
    = -\dfrac{1}{2} (\bar{\chi}_L\gamma_{\mu}\chi_L)(\bar{d}_R\gamma^{\mu}d_R)\;,
\end{align} 
and the relation for Majorana fermion 
\begin{align}
\bar\chi \gamma^\mu \chi = 0\;,
\end{align}
we can obtain the following effective Lagrangian~\cite{Jungman:1995df, Hisano:2010ct, Mohan:2019zrk, Arcadi:2023imv}:
\begin{align}
    \mathcal{L}_\chi = \sum_{i=V,A,S,D} C_i \mathcal{O}_i\;,
\end{align}
where the effective operators are 
\begin{align}
  \mathcal{O}_{V}&=(\bar{\chi}\gamma^{\mu}\gamma^5\chi)(\bar{d}\gamma_{\mu} d)\;, &
    \mathcal{O}_A&=(\bar{\chi}\gamma^{\mu}\gamma^5 \chi)(\bar{d}\gamma_{\mu}\gamma^5 d)\;,\nonumber \\
    \mathcal O_S &= m_d \bar \chi \chi \bar{d} d \;,&
    \mathcal{O}_{D}&=\left[\bar{\chi}i(\partial^{\{\mu}\gamma^{\nu\}}) \chi\right]\left[\bar{d}(\gamma^{\{\mu}iD_{-}^{\nu \}}-\frac{g^{\mu\nu}}{4}i\slashed{D}_{-})d\right]\;.
\end{align}
Here, the notation of the operator $\mathcal{O}_D$ follows Refs.~\cite{Hisano:2010ct,Hill:2014yka},
\begin{align}
    A^{\{\mu}B^{\nu\}}&=(A^{\mu}B^{\nu}+A^{\nu}B^{\mu})/2\;,\nonumber\\
    D_{\pm}^{\mu}=D^{\mu}\pm \overleftarrow{D}^\mu\;,&\quad
    D_{\mu}=\partial_{\mu}-ig_sG_{\mu}^aT^a-ieQA_{\mu} \;.
\end{align}
The Wilson coefficients are given by
\begin{align}
\label{eq:Wilson}
    C_V &= C_A=\frac{f_{d\chi}^2}{8 (m_{\phi}^2-m_{\chi}^2)}\;,\nonumber\\
    C_S&=\dfrac{f_{d\chi}^2m_{\chi}}{16 (m_{\phi}^2-m_{\chi}^2)^2}\;, \quad C_D=\frac{f_{d\chi}^2}{8 (m_{\phi}^2-m_{\chi}^2)^2}\;.
\end{align}

Based on whether the DM-nucleus interactions depend on the spin of nucleus or not, the DM-nucleus scattering is classified as spin-dependent (SD) and spin-independent (SI) scatterings, respectively. The latter is typically proportional to $A^2$, where $A$ is the mass number of nucleus. 
Table~\ref{tab:DM1} summarizes the suppression factors of DM-{\it nucleon} cross sections for the above operators. Here, $\Vec{v}$ is  the DM-nucleon relative velocity, $\Vec{q}$ denotes the momentum transfer, and the transverse relative velocity is defined as $\Vec{v}^{\perp} \equiv \Vec{v}-\Vec{q}/(2\mu_N) $, where $\mu_N \equiv m_\chi m_N/(m_\chi + m_N) $ denotes the DM-nucleon reduced mass.  

\begin{table}[H]
\tabcolsep=4pt
\begin{center}
\renewcommand{\arraystretch}{2.0}
\renewcommand{\arraystretch}{2.0}
\begin{tabular}{c|c|c|c}
\hline
\hline
  & $\mathcal{O}_{V}$ &  $\mathcal{O}_{A}$ & $\mathcal{O}_{S}$ \& $\mathcal{O}_{D}$ \\ \hline
$\sigma_{\text{SI}}$& $ v^{\perp 2}$  &$q^2 v^{\perp 2}$ & $m_{\chi}^2m_N^2/m_{\phi}^4$ \\ \hline
$\sigma_{\text{SD}}$ &  $q^2$  & 1 &$-$ \\ \hline
\hline
\end{tabular}
\caption{
 The suppression factors of the DM-nucleon SI and SD cross sections for the operators $\mathcal{O}_V$, $\mathcal{O}_A$, $\mathcal{O}_S$ and $\mathcal{O}_D$~\cite{Kumar:2013iva}. The factor of 1 indicates no suppression. 
 } 
\label{tab:DM1}
\end{center}
\end{table}

The DM-nucleon SI and SD cross sections for the operators $\mathcal{O}_V$ and $\mathcal{O}_A$ depend on the kinematic suppression factors, which agree with the earlier studies~\cite{Bai:2013iqa,DiFranzo:2013vra,An:2013xka, Mohan:2019zrk}. On the contrary, the SI cross sections for the operators $\mathcal{O}_S$ and $\mathcal{O}_D$ are not kinematically suppressed, but suppressed by $m_{\chi}^2m_N^2/m_{\phi}^4$~\cite{Jungman:1995df, Hisano:2010ct, Mohan:2019zrk, Arcadi:2023imv}. This is because $\mathcal{O}_S$ and $\mathcal{O}_D$ are obtained by expanding the propagator of $\phi_d$ at next-to-leading order. Their contributions to the SD cross section are not displayed as they are further suppressed by the DM velocity and momentum transfer~\cite{Kumar:2013iva}.

\begin{figure}[H]
	\centering
    \includegraphics[width=0.55\linewidth]{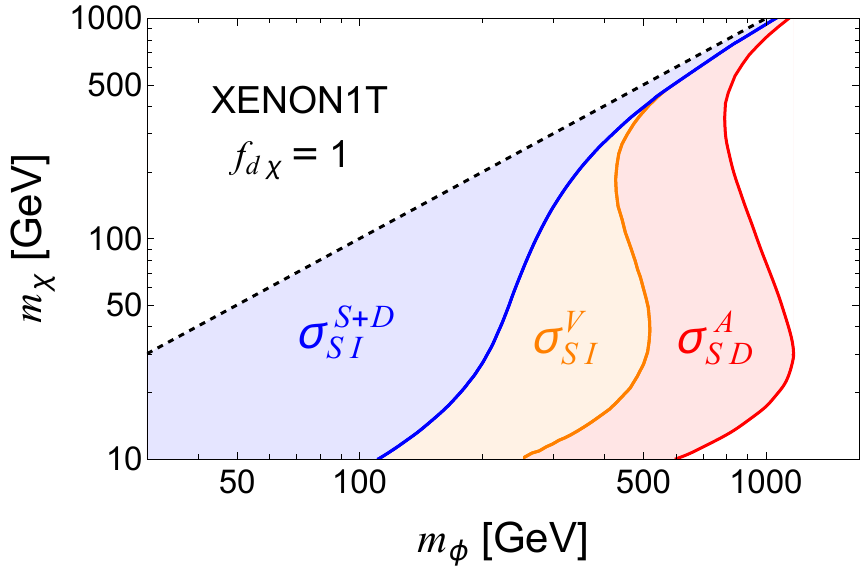}
    \caption{The exclusion limits of $m_\chi$ and $m_\phi$ for $\mathcal{O}_V$, $\mathcal{O}_A$, and a combination of 
    $\mathcal{O}_S$ and $\mathcal{O}_D$ by the XENON1T experiment with a 1 ton-year exposure \cite{XENON:2018voc, XENON:2019rxp} under the assumption of $f_{d\chi}=1$. The DM-nucleon SD cross section is attributed to the operator $\mathcal{O}_A$, while the SI cross sections from $\mathcal{O}_V$ and the combination of $\mathcal{O}_S$ and $\mathcal{O}_D$ are analyzed separately. }
 \label{fig:three operators}	
\end{figure}

We first consider the SI contributions from these operators. From Eq.~\eqref{eq:Wilson}, the Wilson coefficients of the operators $\mathcal{O}_V$ and $\mathcal{O}_A$ are the same, so we readily obtain that the contribution of $\mathcal{O}_V$, which is suppressed by $v^{\perp 2} \sim 10^{-6}$, is larger that that of $\mathcal{O}_A$.
To derive the exclusion limit for the operator $\mathcal{O}_V$, we evaluate the non-relativistic events generated in the XENON1T experiment~\cite{XENON:2018voc} with an exposure of $w = 1.0 \ \text{ton-year}$ using {\tt DirectDM}~\cite{Bishara:2017nnn} and {\tt DMFormFactor}~\cite{Anand:2013yka}. The 90\% confidence level (C.L.) constraint is determined under the assumption of 7 signal events after accounting for SM backgrounds~\cite{Kang:2018rad, Liang:2024tef}. Our result agrees with the EFT analysis by the XENON Collaboration~\cite{XENON:2022avm}. The exclusion region for $m_{\chi}$ and $m_{\phi}$ is depicted in orange color in Fig.~\ref{fig:three operators}.

The suppression factor for the contributions to the SI cross section from the operators $\mathcal{O}_{S}$ and $\mathcal{O}_{D}$ is $m^2_\chi m^2_N/m_{\phi}^4 \sim 10^{-6} \times (m_{\chi}/m_{\phi})^2$. Following Refs.~\cite{Jungman:1995df, Hisano:2010ct, Mohan:2019zrk, Arcadi:2023imv, Hill:2014yka}, we express the contributions to the SI cross section as a combination of the Wilson coefficients of $\mathcal{O}_S$ and $\mathcal{O}_D$.
In Fig.~\ref{fig:three operators}, we present the exclusion of $m_\chi$ and $m_\phi$ using the upper limit on the DM-nucleon SI cross sections by the XENON1T experiment~\cite{XENON:2018voc} with a 1 ton-year exposure.  We find that the contribution from $\mathcal{O}_V$ dominates over that from the combination of $\mathcal{O}_{S}$ and $\mathcal{O}_{D}$ within our parameter space.

On the other hand, 
the leading contribution to the SD cross section comes from the operator $\mathcal{O}_A$, as clearly shown in Table~\ref{tab:DM1}, where the contributions from $\mathcal{O}_S$ and $\mathcal{O}_D$ are highly suppressed and not presented. The SD DM-nucleon cross section for $\mathcal{O}_A$ is expressed as~\cite{Bai:2013iqa,DiFranzo:2013vra,An:2013xka,Mohan:2019zrk}
\begin{align}
\label{SDA}  
\sigma_{\mathrm{SD}}^{A}=\frac{3 f_{d\chi}^4 \mu_{n}^2(\Delta_{d}^n)^2}{16 \pi(m_{\phi}^2-m_\chi^2)^2}\;,  
\end{align}
where $\Delta_d^n = 0.842 \pm 0.012$~\cite{Belanger:2008sj} and $\mu_n$ denotes the DM-neutron reduced mass. Here, we only consider the limit on the DM-neutron cross section, which is much stronger than that on the DM-proton cross section for the XENON1T~\cite{XENON:2019rxp}, PandaX-4T~\cite{PandaX:2022xas}, and LZ~\cite{LZ:2022lsv} experiments. Using Eq.~\eqref{SDA} and the upper limit on the DM-nucleon SD cross section given by the XENON1T experiment~\cite{XENON:2019rxp} with a 1 ton-year exposure, we derive the exclusion of $m_{\chi}$ and $m_{\phi}$ for the operator $\mathcal{O}_A$, which is shown as the red region in Fig.~\ref{fig:three operators}. It is evident that in our case, the exclusion limit from DM direct detection is dominated by the SD interactions associated with the operator $\mathcal{O}_A$.

\begin{figure}[!htb]
	\centering
\includegraphics[width=0.5\linewidth]{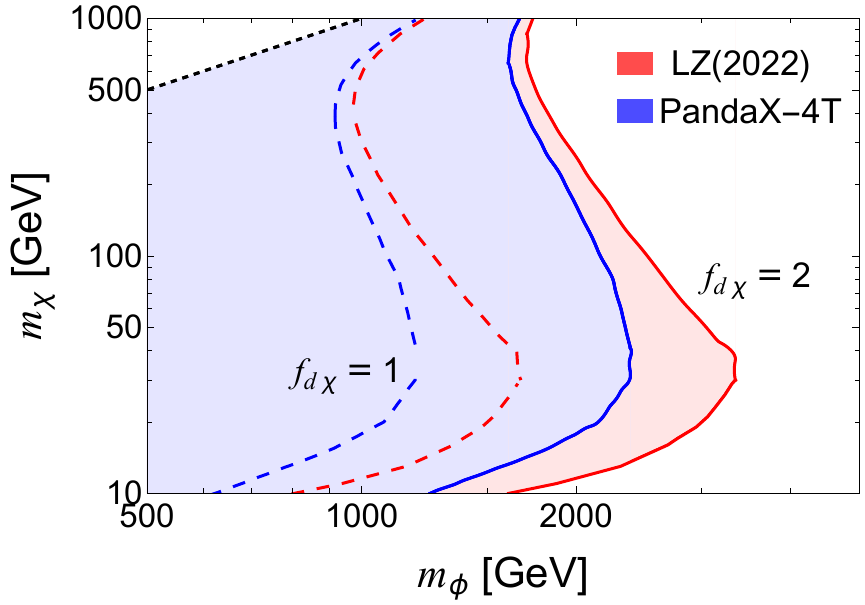}
	\caption{The exclusion regions of $m_\chi$ and $m_\phi$ for constraints on SD cross sections from the LZ~\cite{LZ:2022lsv} and PandaX-4T~\cite{PandaX:2022xas} experiments shown in red and blue colors, respectively, for the assumption of the coupling $f_{d\chi}=2$ (solid curve) or 1 (dashed curve).
 }
 \label{fig:dd}	
\end{figure}

In Fig.~\ref{fig:dd},  we show the exclusion limits on the DM mass $m_\chi$ and colored mediator mass $m_\phi$ from the constraints on the SD cross sections measured in the LUX-ZEPLIN (LZ) experiment~\cite{LZ:2022lsv} and the PandaX-4T experiment~\cite{PandaX:2022xas} assuming coupling values $f_{d\chi} = 1$ or 2. Our result for $f_{d\chi} = 1$ using the limit by the LZ experiment agrees with Ref.~\cite{Arcadi:2023imv}. It is important to mention that we assume the relic density of $\chi$ as the observed total DM relic density. If the DM relic density depends on a given $f_{\chi S}$, the limits could be notably weaker, as we will see in Sec.~\ref{sec:result}.

It is noted that direct detection can also occur by exchanging photons at the one-loop level,  which leads to much weaker direct detection signals compared to the contribution from exchanging $\phi_d$ at the tree level; see Appendix~\ref{app:anapole} for details.

\section{Collider searches}
\label{collider Search}

The DM $\chi$ can interact with SM particles via $F$, $S$ and $\phi_d$, resulting in detectable DM signatures at collider experiments. We focus on DM searches at the LHC, where the presence of missing energy is a typical signature indicating the possible existence of DM particles. 
Assuming the mass hierarchy $m_F \geq m_\phi > m_S > m_\chi$,
and the decay branching ratios of $S$ and $\phi_d$ are determined as
\begin{align}
{\rm BR}(S^\pm \to \chi \mu^\pm)=1\;, \quad
{\rm BR}(\phi_d^{\pm 1/3} \to \chi d)=1\;.
\end{align}
The particle $F$ can decay to leptons or jet via the processes $F\to \phi_d j (S^\pm \ell^\mp)$ or $F^\pm \to \phi_d j (S^\pm \nu)$, where 
$j$ represents any first-generation quark. The most relevant collider search for DM involves these channels: (1) leptons$+\slashed E_T$ and (2) jet(s)$+\slashed E_T$. 

\subsection{Leptons\tf{$+\slashed E_T$}{+MET}}
The possible processes for leptons$+\slashed E_T$ are as follows:
    \begin{itemize}
        \item SL1: $pp\to S^+ S^-$, $S^\pm \to \mu^\pm \chi$;
        \item SL2: $pp\to F^+ F^-$, $F^+ \to S^+ \nu_{\ell}$, $F^- \to S^- \bar{\nu}_{\ell}$, $S^\pm \to \mu^\pm \chi$;
        \item SL3: $pp\to F^0 \bar{F}^0$, $F^0 \to S^+ \ell^-$, $\bar{F}^0 \to S^- {\ell}^+$, $S^\pm \to \mu^\pm \chi$, $\ell = e$ or $\mu$.
    \end{itemize}
\begin{figure}[!htb]
    \centering
    \includegraphics[scale=0.35]{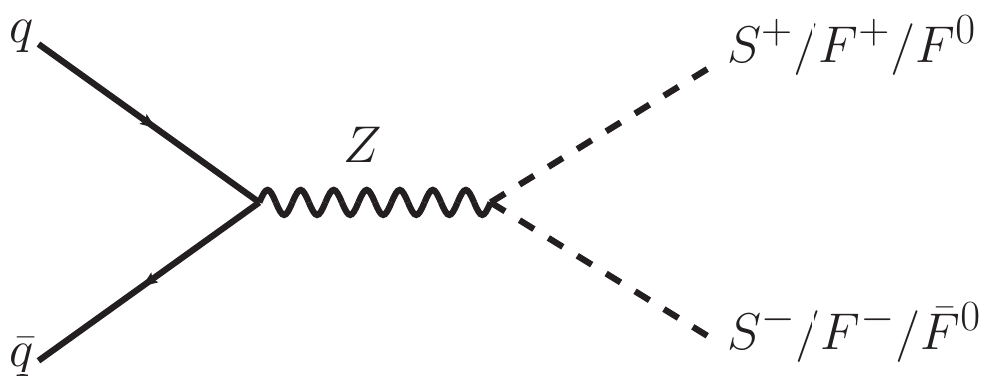}\qquad\qquad
    \includegraphics[scale=0.35]{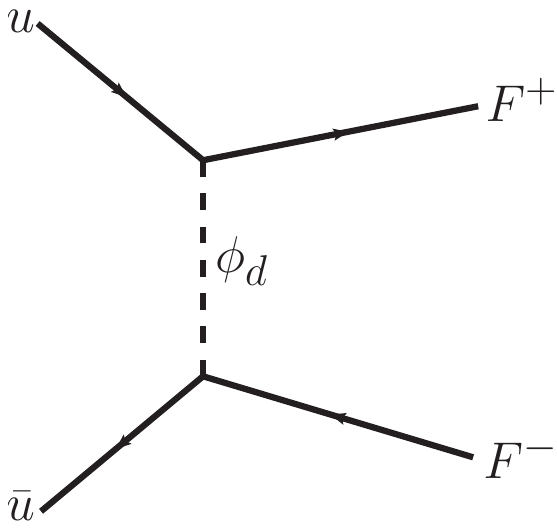}\qquad
    \includegraphics[scale=0.35]{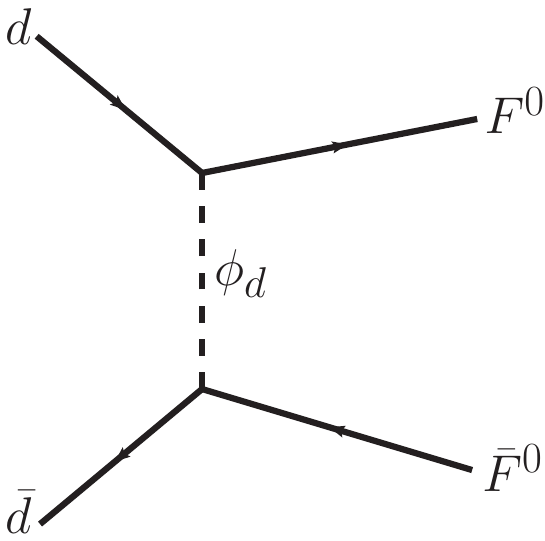}\qquad
      \caption{Representative diagrams for pair production of $S^{\pm}$, $F^{\pm}$ and $ F^0$. In the left panel, $S^+ S^-$ and $F^+ F^-$ can also be produced with the $s$-channel off-shell photon.}
    \label{fig:collider-Spair}
\end{figure}

\begin{figure}[!htb]
    \centering
    \includegraphics[width=0.55\linewidth]{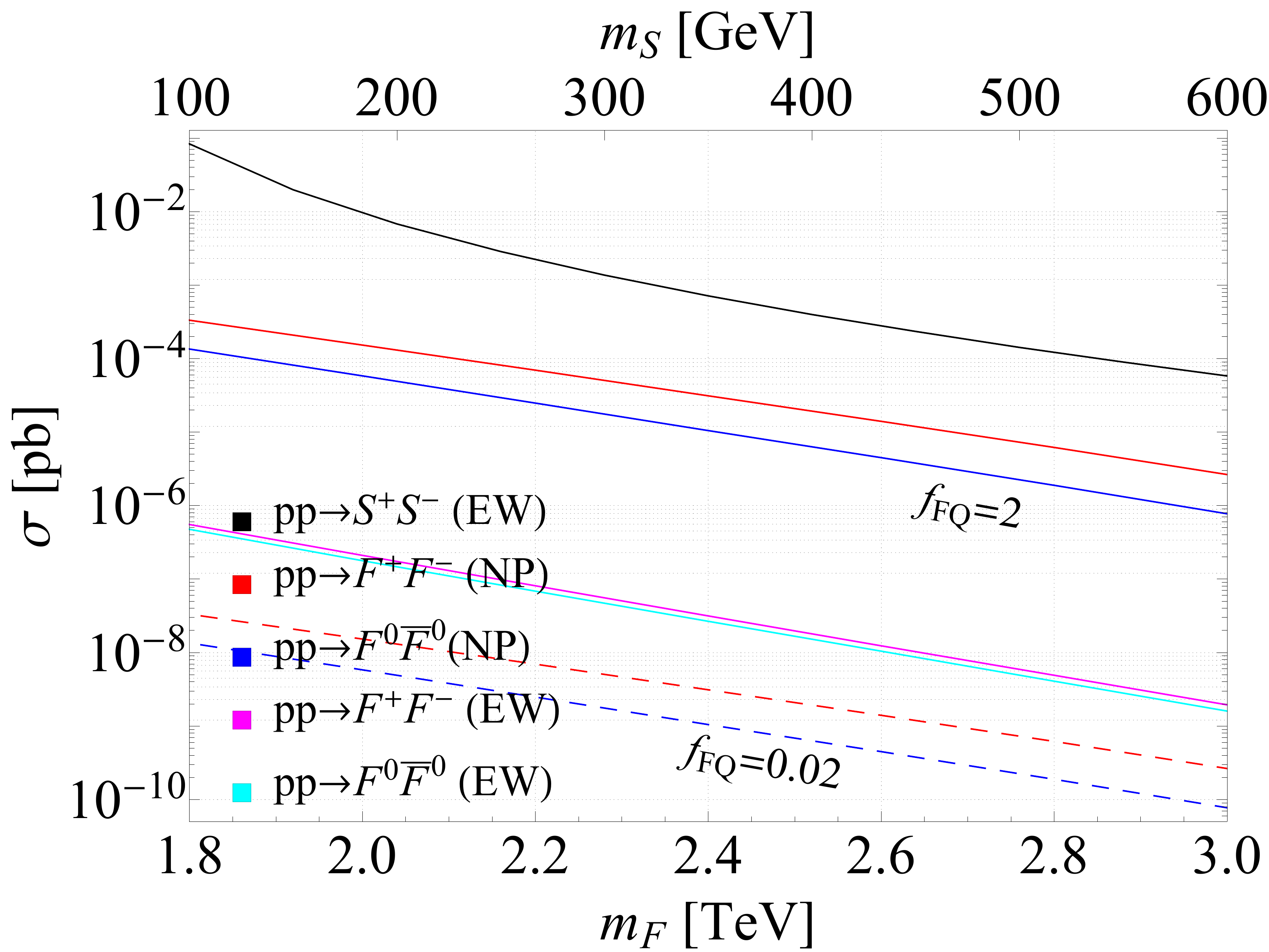}
    \caption{Production cross sections for the production of $F^+ F^-$ and $S^+ S^-$ at the 13~TeV LHC with the masses of $F$ and $S$ depicted in the lower and upper axes, respectively, for the assumption of $m_\phi = 2.5\tev$. 
    The new physics coupling $f_{FQ}$ is assumed to be 2 (solid red, solid blue) or 0.02 (dashed red, dashed blue).
    }
    \label{fig:leptonsMET_xsec}
\end{figure}

The corresponding Feynman diagrams for the production of $S^+ S^-$, $F^+ F^-$ and $F^0 \bar F^0$ are illustrated in Fig.~\ref{fig:collider-Spair}, which depend on the electroweak or new physics couplings. The cross sections for production of $F^+ F^-$ and $F^0 \bar F^0$ in the middle and right panels
 are proportional to $f_{FQ}^4$ and insensitive to the mass $m_\phi$.
In our study, we consider the mass ranges $m_F\in[2,3]$~TeV and $m_S\in[100, 500]$~GeV, and assume $m_\phi = 2.5\tev$ and the coupling $f_{FQ}=2$ or 0.02.
The cross sections  at the 13~TeV LHC are compared in Fig.~\ref{fig:leptonsMET_xsec}. It is shown that the cross section of
$S^+S^-$ pair production via electroweak processes, as depicted in the left panel of Fig.~\ref{fig:collider-Spair}, is larger than the others within our specified mass ranges.
Therefore, we focus exclusively on the $S^+S^-$ signal (denoted as SL1) in the recast of dilepton$+\slashed E_T$ searches.

\begin{figure}[!htb]
    \centering
        \includegraphics[width=0.55\linewidth]{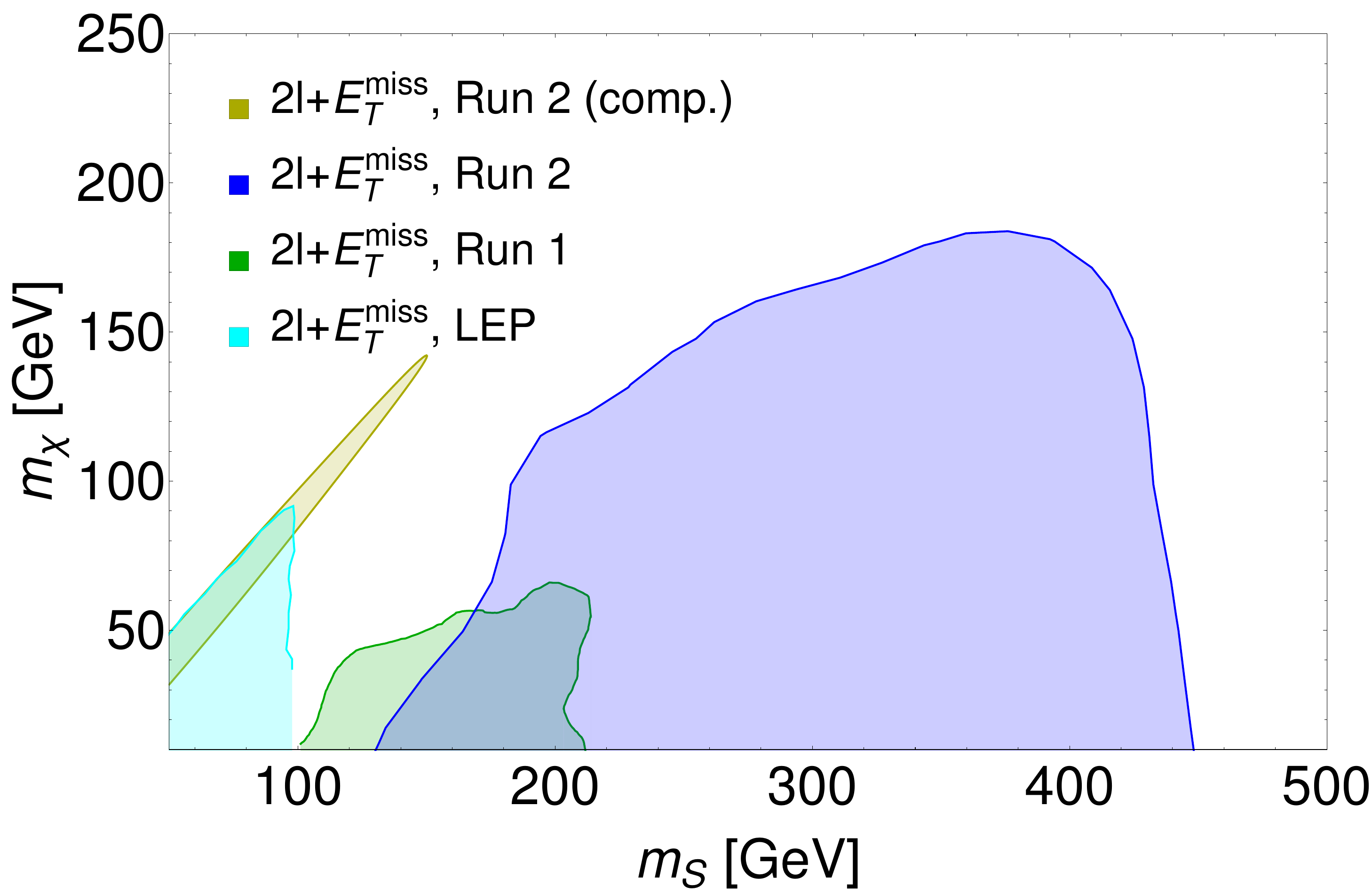}
        \caption{Shaded regions in blue, green, and dark yellow colors excluded by the dilepton$+\slashed E_T$ searches at the LHC Run 1 ($8\tev$, $20.3\fbi$)~\cite{ATLAS:2014zve}, Run 2 $(139\fbi)$ in the scenarios of $m_\chi < m_S$~\cite{ATLAS:2019lff} and $m_S - m_\chi \ll m_S$ (compressed)~\cite{ATLAS:2019lng}, respectively. The cyan region excluded by the LEP is taken from Ref.~\cite{Liu:2021mhn}.
        }
    \label{fig:dileptonMET}
\end{figure}

Since the scalar $S^\pm$ interacts with the right-handed charged leptons, which ensembles the right-handed slepton~\cite{Fuks:2013lya}, and decays totally into $\mu^\pm \chi$, one can read off the exclusion limits on the masses of right-handed slepton (smuon) and neutralino, and reinterpret them as the constraints on
the masses of $m_S$ and $m_\chi$. 
In Fig.~\ref{fig:dileptonMET}, we show the excluded regions by the most stringent searches at the LHC Run 2~\cite{ATLAS:2019lff,ATLAS:2019lng} and earlier 8~TeV LHC~\cite{ATLAS:2014zve}. 

Besides the dilepton$+\slashed E_T$ processes, the cascade decays of $F^0/\bar F^0$ and $S^\pm$ can give rise to the multi-lepton$+\slashed E_T$ process SL3. For $m_F = 3\tev$, the cross section for the production of $F^0 \bar F^0$ is $7.7\times 10^{-7}{~\rm pb}$, for which less than 3 signal events are expected even with the integrated luminosity of $3\abi$. Thus there is no constraint from the multi-lepton$+\slashed E_T$ searches.

\subsection{Jet(s)\tf{$+\slashed E_T$}{+MET}}
Depending on the number of jets at the parton level, the possible process for jet(s)$+\slashed E_T$ are following:
    \begin{itemize}
        \item SJ1: $pp\to \chi \bar\chi j$ (3-body), $j=g$, $d$ or $\bar d$;
        \item SJ2: $pp\to \phi_d^{\pm1/3} \chi$, $\phi_d^{\pm1/3} \to \bar\chi j$;
        \item SJ3: $pp\to \phi_d^{+1/3} \phi_d^{-1/3}$, $\phi_d^{\pm1/3} \to \chi j$;
    \end{itemize}

\begin{figure}[!htb]
\centering
    \includegraphics[scale=0.55]{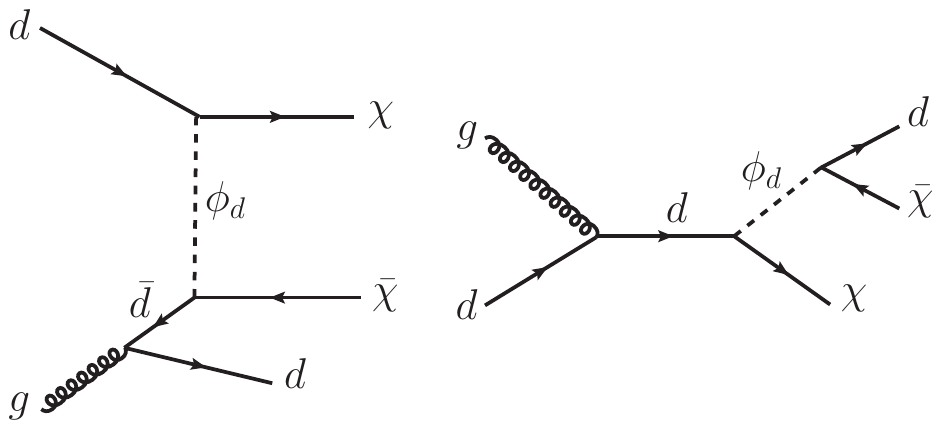}
    \caption{Representative diagrams for the signals SJ1 (left) and SJ2 (right) in the monojet$+\slashed E_T$ channel.  }
    \label{fig:monojet+MET}
\end{figure}

\begin{figure}[!htb]
    \centering
    \includegraphics[width=0.55\linewidth]{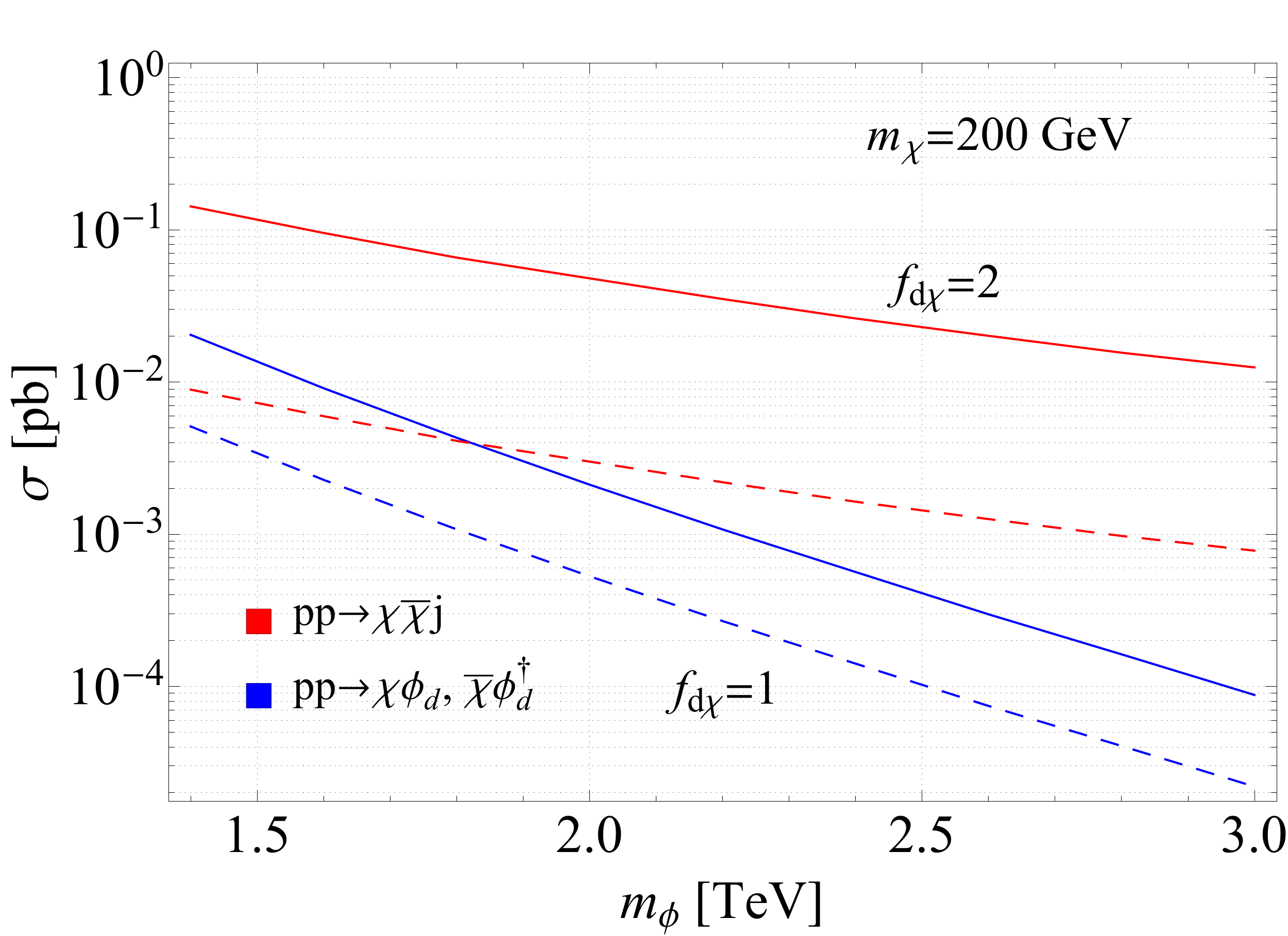}
    \caption{Cross sections for the production of signals SJ1 (red) and SJ2 (blue) in the monojet$+\slashed E_T$ channel at the 13~TeV LHC for the assumptions of $f_{d\chi}=2$ (solid) and $1$ (dashed), and $m_\chi = 200\gev$.
   }
    \label{fig:monojetMET_xsec}
\end{figure}

The signal processes SJ1 and SJ2 for DM production at the LHC are categorized as monojet$+\slashed E_T$ channel, while the signal process SJ3 is categorized as dijet$+\slashed E_T$ channel. The representative diagrams for monojet$+\slashed E_T$ are illustrated in Fig.~\ref{fig:monojet+MET}.
The cross sections of $pp\to \chi \bar \chi j$ and $pp\to \phi_d \chi$ scale with $f_{d\chi}^4$ and $f_{d\chi}^2$, respectively. In Fig.~\ref{fig:monojetMET_xsec}, we illustrate these cross sections with the new physics coupling $f_{d\chi}=2$ (thick lines) or 1(dashed lines) at the 13~TeV LHC.
It is evident that, without any selection cuts applied, the cross section of SJ1 is much larger than SJ2, especially for a larger $m_\phi$.

\begin{figure}[!htb]
    \centering
    \includegraphics[scale=0.45]{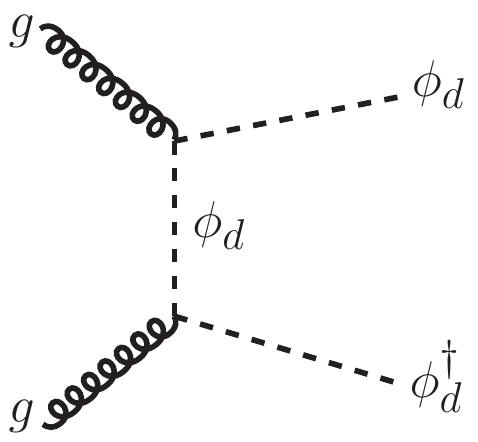}\qquad
    \includegraphics[scale=0.45]{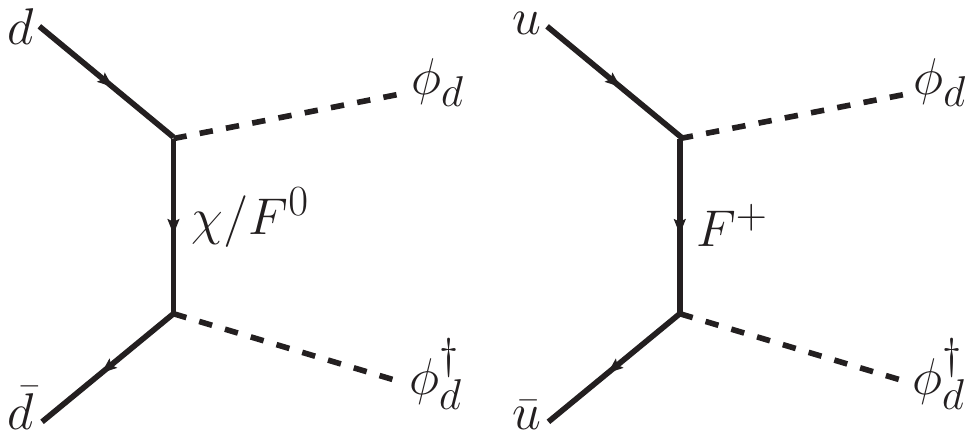}
    \caption{Feynman diagrams for the dijet$+\slashed E_T$ processes. Left panel: QCD production. Right two panels: new physics production. }
    \label{fig:dijet+MET}
\end{figure}

\begin{figure}[!htb]
    \centering
    \includegraphics[width=0.55\linewidth]{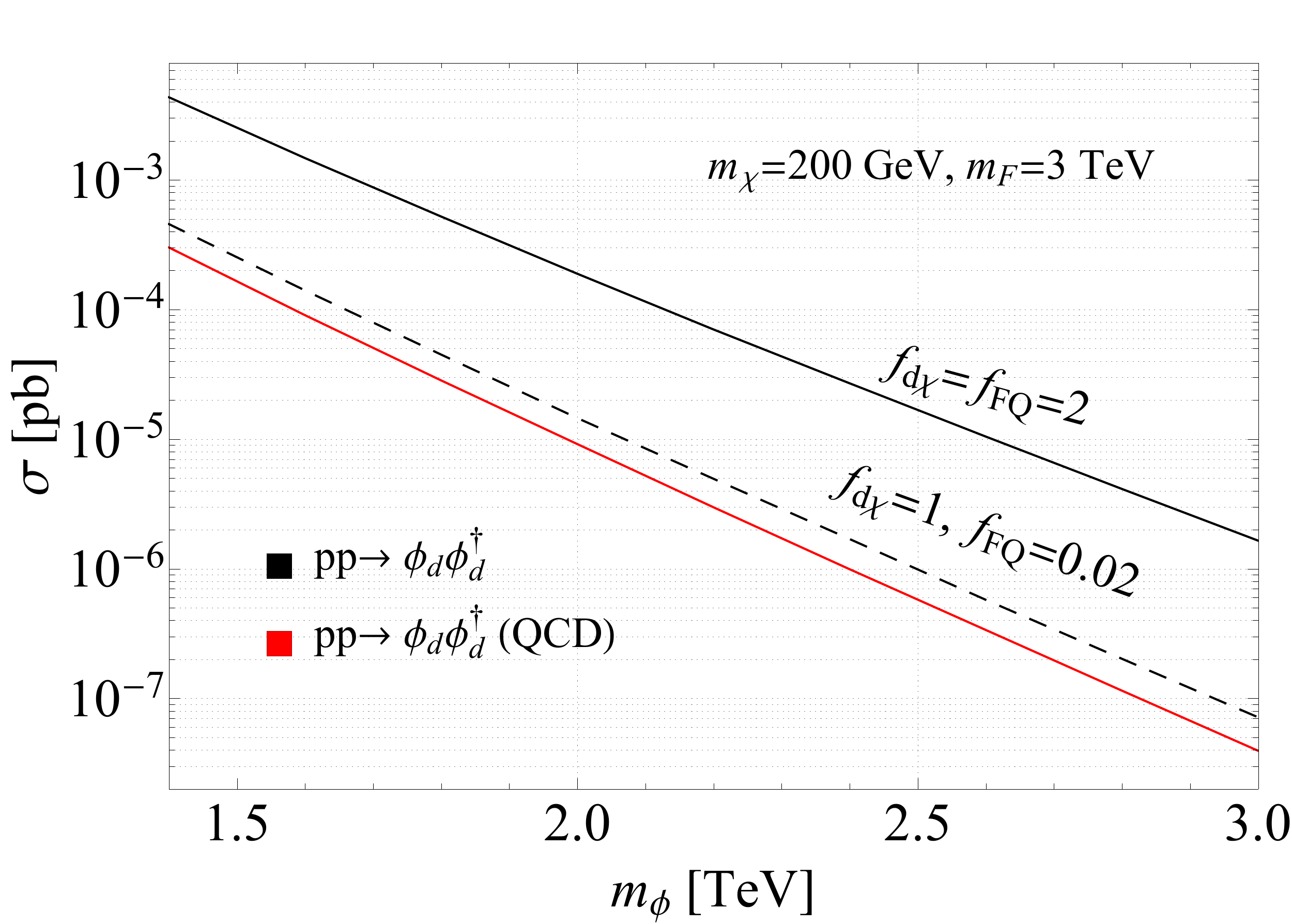}
    \caption{Cross sections for the production of $\phi_d \phi_d^\dagger $ at the 13~TeV LHC for the assumptions of $f_{d\chi}=f_{FQ}=2$ (black solid) or $f_{d\chi}=1$, $f_{FQ}=0.02$ (black dashed), $m_\chi=200\gev$, and $m_F=3\tev$. The red line corresponds to the QCD production. 
    }
    \label{fig:dijetMET_xsec}
\end{figure}

In the dijet$+\slashed E_T$ channel with two jets at the parton level, the DM $\chi$ is produced from the decay of colored mediator $\phi_d$. In Fig.~\ref{fig:dijet+MET}, we show the representative diagrams for the production via QCD or new physics interactions, respectively. 
The cross section for the QCD production of $\phi_d \phi_d^\dagger $ (red line), and benchmark cross sections for the new physics production at the 13~TeV LHC under the assumptions of $m_\chi=200\gev$, $m_F=3\tev$, and $f_{d\chi}=f_{FQ}=2$ (black thick line),  or $f_{d\chi}=1$, $f_{FQ}=0.02$ (black dashed line) are illustrated in Fig.~\ref{fig:dijetMET_xsec}. We find that the total production cross section is significantly enhanced for substantial new physics couplings, which agrees with the findings in Ref.~\cite{Bai:2013iqa}. In our parameter space of interest, the cross section of new physics production always dominates over that of QCD production.

At the detector level, more jets would emerge from the signal processes discussed, even if only one or two jets are present at the parton level. Therefore, a detailed analysis must include detector-level simulations. Next, we separate the analysis into monojet and dijet searches.

\subsubsection{Monojet search}

\begin{figure}[!hbtp]
    \centering
        \includegraphics[width=0.55\linewidth]{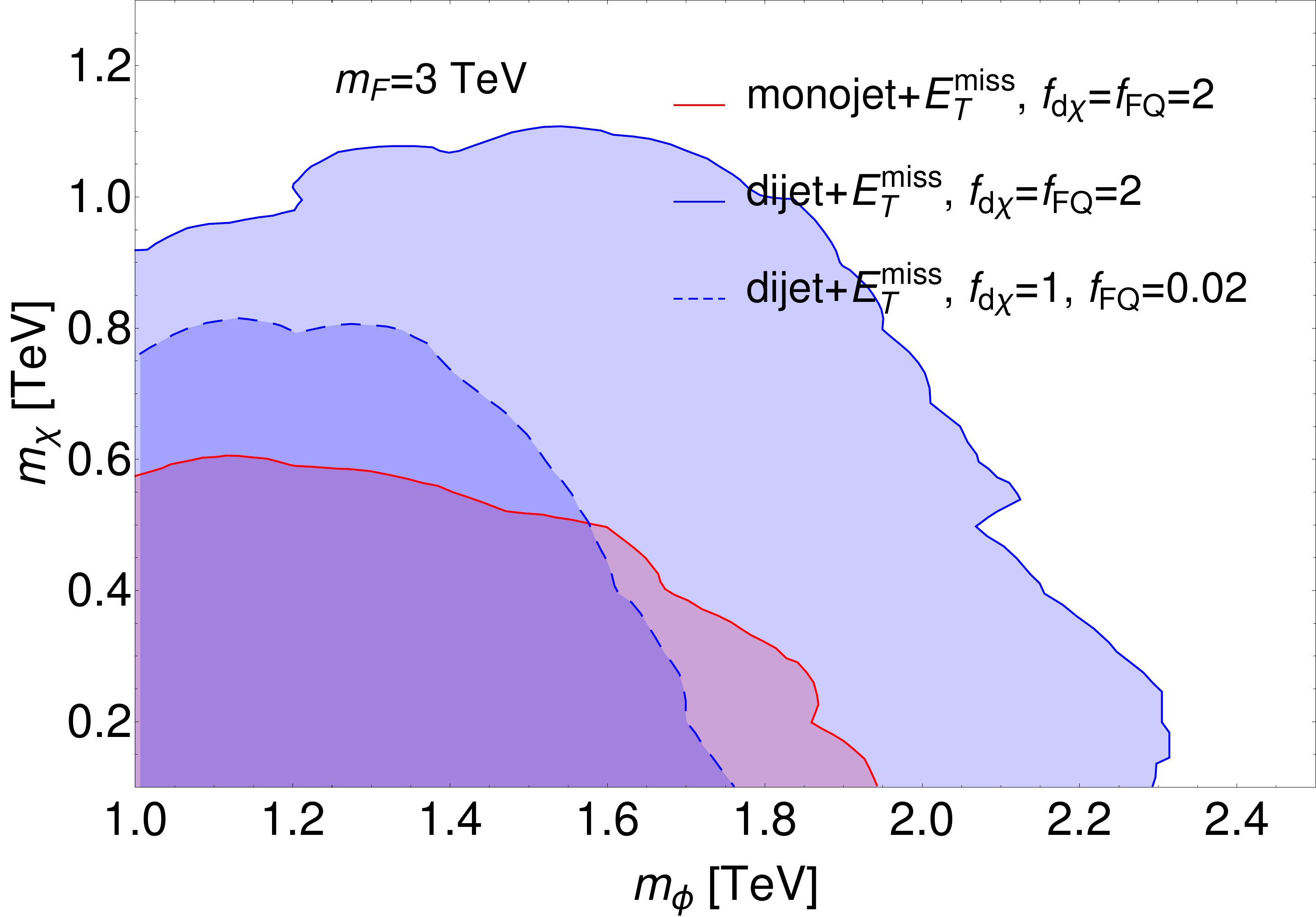} 
        \caption{Shaded regions in red and blue colors excluded by  the monojet$+\slashed E_T$~\cite{ATLAS:2021kxv} and dijet$+\slashed E_T$~\cite{ATLAS:2021kxv} searches at the LHC Run 2 $(139\fbi)$, respectively, for the assumptions of $f_{d\chi} = f_{FQ}=2$ (solid curves) or $f_{d\chi}=1$, $f_{FQ}=0.02$ (dashed curve), with $m_F=3\tev$ being fixed.
        }
    \label{fig:exlusion_jetMET}
\end{figure}

In the jet(s)$+\slashed E_T$ searches, it is generally essential to consider the processes SJ1, SJ2, and SJ3 simultaneously, including additional jets from the initial state radiation (ISR). However, for monojet searches, which require one hard jet, we only need to consider SJ1 and SJ2, since the parton-level cross section of SJ3 is much smaller.
We generate signal events using \texttt{MadGraph5\_aMC@NLO}~\cite{Alwall:2014hca}, which are then passed to \texttt{Pythia8}~\cite{Sjostrand:2014zea} for parton shower, and \texttt{Delphes3}~\cite{deFavereau:2013fsa} for detector simulation, respectively. Following the most stringent monojet$+\slashed E_T$ search~\cite{ATLAS:2021kxv} at the LHC Run 2, we apply the selection cuts as follows:
\begin{itemize}
\item $\slashed E_T >200~\text{GeV}$, leading jet $p_T> 150$~GeV and $|\eta| < 2.4$;
\item  up to four jets with $p_T> 30$~GeV and $|\eta| < 2.8$;
\item   $|\Delta \phi( \text{jet}, \slashed {\bf p}_T)|> 0.4$ ($0.6$) for $\slashed E_T >250\gev$ ($< 250\gev$);
\item veto of electron, muon, $\tau$-lepton or photon.
\end{itemize}
We further refine our event selection according to 13 inclusive signal regions. By combining model-independent constraints on the observed signal cross section $\sigma$ for each of these regions, as outlined in Ref.~\cite{ATLAS:2021kxv}, we derive a $95\%$ C.L. limit on our parameter space of $m_\phi$ and $m_\chi$. The excluded region of $m_\phi$ and $m_\chi$ is depicted by the red shaded area in Fig.~\ref{fig:exlusion_jetMET} assuming the new physics coupling $f_{d\chi} = f_{FQ}=2$, for which the monojet$+\slashed E_T$ search at the LHC Run 2 sets the lower limit $m_\phi \geq 1.93\tev$. The exclusion for $f_{d\chi}=1$, $f_{FQ}=0.02$ falls below $1\tev$ and is not depicted. This result agrees with the reinterpretation in Ref.~\cite{Mohan:2019zrk} of the monojet$+\slashed E_T$ search at the LHC using the data of $36.1\fbi$~\cite{ATLAS:2017bfj} assuming that the signal significance approximately scales as the square root of the integrated luminosity.

\subsubsection{Dijet search}

To recast the dijet$+\slashed E_T$ search, we consider all of the processes SJ1, SJ2, and SJ3, including their ISR, to generate detector-level events. We implement the selection criteria outlined in Ref.~\cite{ATLAS:2020syg}. The signal events are first selected with the following criteria:
\begin{itemize}
\item $\slashed E_T >300~\text{GeV}$, leading jet with $p_T(j_1)> 200$~GeV and sub-leading jet with $p_T(j_2)> 50$~GeV, and $|\Delta \phi( j_{1,2,(3)}, \slashed {\bf p}_T)|> 0.2$ ;
\item  $m_{\text{eff}}> 800$ GeV;
\item veto of electron (muon) with $p_T>6 (7)$ GeV,
\end{itemize}
where $m_{\text{eff}}$ is defined by the scalar sum of $\slashed E_T$ and 
transverse momentum of jets with $p_T>50$~GeV. We further apply the selection criteria of model-independent search SR2j-1600, SR2j-2200, and SR2j-2800 that are listed in Ref.~\cite{ATLAS:2020syg}. We then derive our constraints from the model-independent limits of the three signal regions given in  Ref.~\cite{ATLAS:2020syg} accordingly.  

In Fig.~\ref{fig:exlusion_jetMET}, we show the constraints from the monojet and dijet searches for $m_F=3\tev$\footnote{For $f_{d\chi}=1$, $f_{FQ}=0.02$, the cross section from the $t$-channel exchange of $F$ is negligible compared to that from the exchange of $\chi$,  thus the total cross section is insensitive to the mass of $F$ in this case.
} with different choices of the new physics couplings. The blue shaded areas with solid and dashed curves correspond to dijet search with $f_{d\chi} = f_{FQ} = 2$ and $f_{d\chi} =1$, $ f_{FQ} = 0.02$, respectively.
The constraints from the dijet search are stricter than those from the monojet search with the same choice of $f_{d\chi}$ and $f_{FQ}$.
We find that the most stringent limits on the colored mediator mass from the jet(s)$+\slashed E_T$ searches in our parameter space are $m_\phi \geq 2.3\tev$ and $1.76\tev$, respectively.

\section{Results and discussion}
\label{sec:result}

We have explored the phenomenological implications of the four-fermion operator $O_{ledq}^{\alpha\beta 11}$ in the simplified model, including DM relic density, direct detection, and collider searches, as well as the model-independent LFC and LFV constraints on the Wilson coefficients. To analyze their complementarities, we examine four benchmark (BM) scenarios detailed in Table~\ref{tab:cases}.

\begin{table}[!htb]
\tabcolsep=4pt
\begin{center}
\renewcommand{\arraystretch}{1.6}
\begin{tabular}{c|c|c|c|c|c|c}
\hline
\hline
 BM &$m_{\phi}$[TeV] & $m_{F}$ [TeV] & $f_{LS}=f_{FQ}$   & $f_{d\chi}$ &  $f_{\chi S}$ & $\Omega_\chi h^2$ \\ \hline
(a) & 2.5 & 3.0 & 2.1  & 2.0  & 2.0  & / \\ \hline
(b) & 2.5 & 3.0 & 2.1   & 2.0  & /  & 0.1199 \\ \hline
(c) & 2.0 & 2.0& $1.41 \times 10^{-2}$   & 1.0 & 1.5 & / \\ \hline
(d) & 2.0 & 2.0 & $1.41 \times 10^{-2}$   & 1.0  & / & 0.1199 \\ \hline
\hline
\end{tabular}
\end{center}
\caption{
Benchmark couplings and masses of new particles. For BM~(b) and BM~(d), the coupling $f_{\chi S}$ is determined by the requirement of DM relic density $\Omega_\chi h^2 = 0.1199$. 
}
\label{tab:cases}
\end{table}

Based on the collider searches and DM direct detection discussed, 
we choose the values: 
\begin{itemize}
    \item BM~(a) or (b): $f_{d\chi}=2$, $m_\phi= 2.5 {~\rm TeV}, m_F=3 {~\rm TeV}$;
    \item BM~(c) or (d): $f_{d\chi}=1$, $m_\phi= m_F= 2 {~\rm TeV}$.
\end{itemize}
As discussed in Sec.~\ref{sec:relic_density}, the DM relic density can be determined by the coupling $f_{\chi S}$ and the mass $m_\chi$, $m_S$. In BM~(a) and BM~(b), the coupling $f_{\chi S}$ is taken as an input parameter.
Given the constraints on the LFC and LFV Wilson coefficients $C_{ledq}^{\alpha\beta 11}/\Lambda^2$, as illustrated in Fig.~\ref{Fig:wilson_contour},
the other couplings are set to be equal, such that their products yield $f_{\rm NP} = 2.05$ and $0.1314$ (cf. Eq.~\eqref{eq:fNP-coupling}), respectively. In BM~(b) and BM~(d), $f_{\chi S}$ is fixed by the observed DM relic density for comparison.

\begin{figure}[!htb]
\centering
\includegraphics[width=0.48\linewidth]{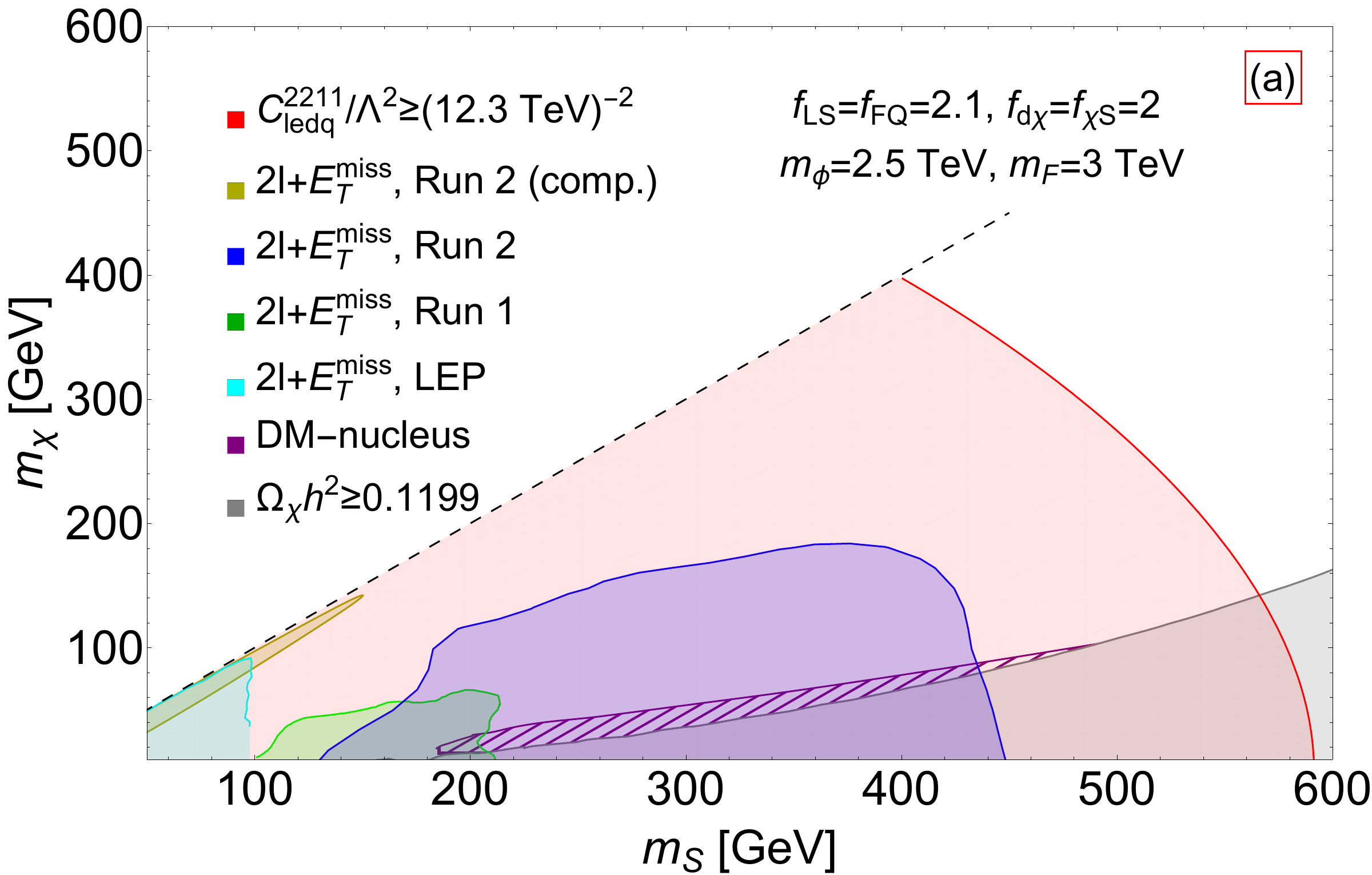}
\includegraphics[width=0.48\linewidth]{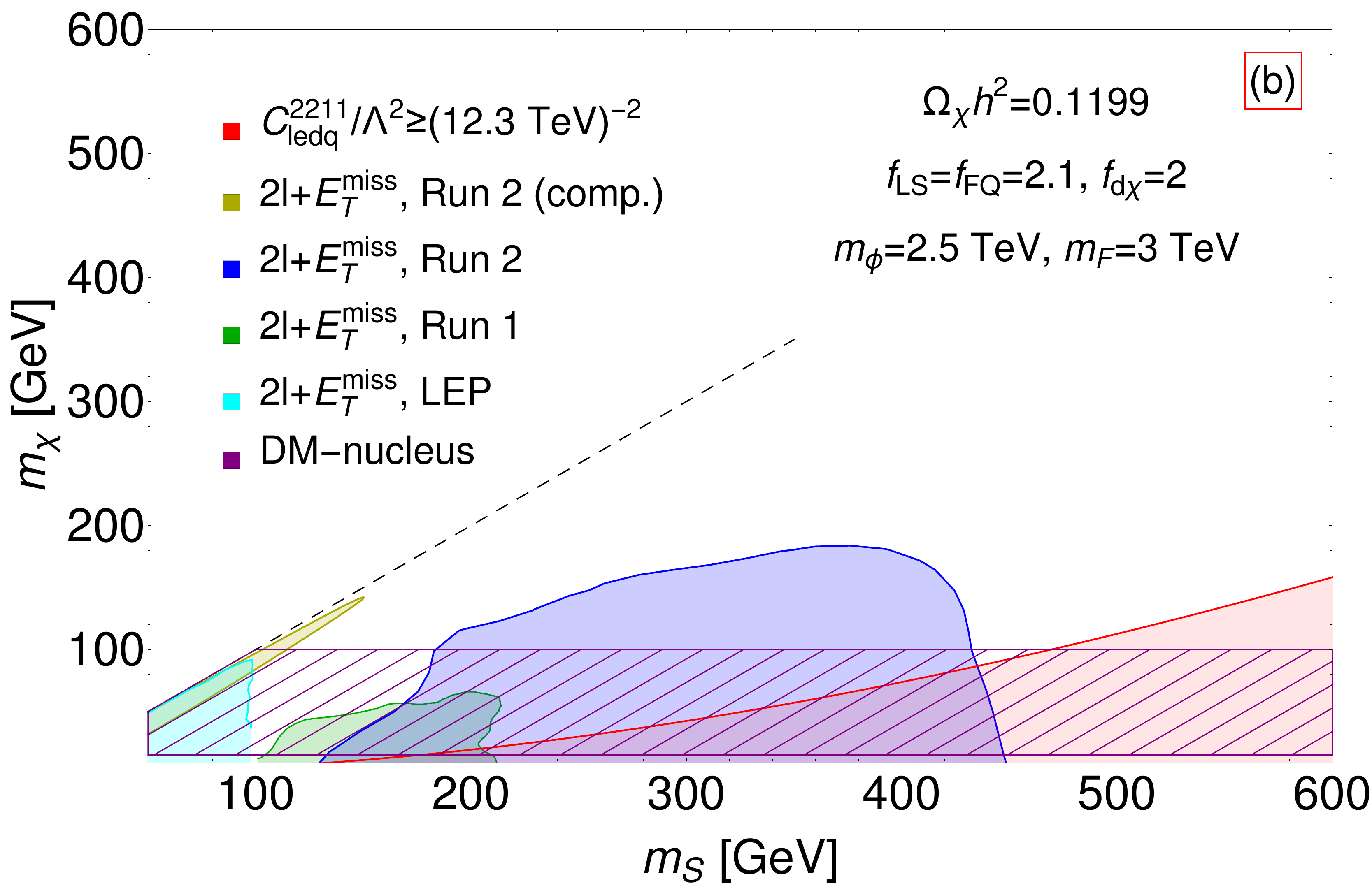}
     \caption{Combined sensitivities to the DM mass $m_\chi$ and the scalar mass $m_S$ for BM~(a) and BM~(b) with $ f_{d\chi} = 2.0$, $f_{LS}=f_{FQ}=2.1$, $m_\phi = 2.5\tev$ and $m_F = 3\tev$. {\bf Left}: $f_{\chi S} =2.0$ and $\Omega_\chi h^2 \leq 0.119$ are required for BM~(a). {\bf Right}: $f_{\chi S}$ is determined by $\Omega_\chi h^2 = 0.119$ for BM~(b).    
    The projected future LFC constraint $C_{ledq}^{2211}/\Lambda^2< (12.3~\text{TeV})^{-2}$ \cite{Du:2020dwr, Du:2021rdg} from the neutrino NSI measurements is expected to exclude the red shade area.
     Constraints from existing dilepton$+\slashed E_T$ searches as the same as those in Fig.~\ref{fig:dileptonMET}. 
     The exclusion resulting from the current constraint from DM direction detection in the LZ experiment~\cite{LZ:2022lsv} is depicted by the purple slash shading region.
     The dashed line represents $m_\chi = m_S$.}
     \label{fig:combined_flavor_conserving} 
\end{figure}

\begin{figure}[!htb]
\centering
\includegraphics[width=0.48\linewidth]{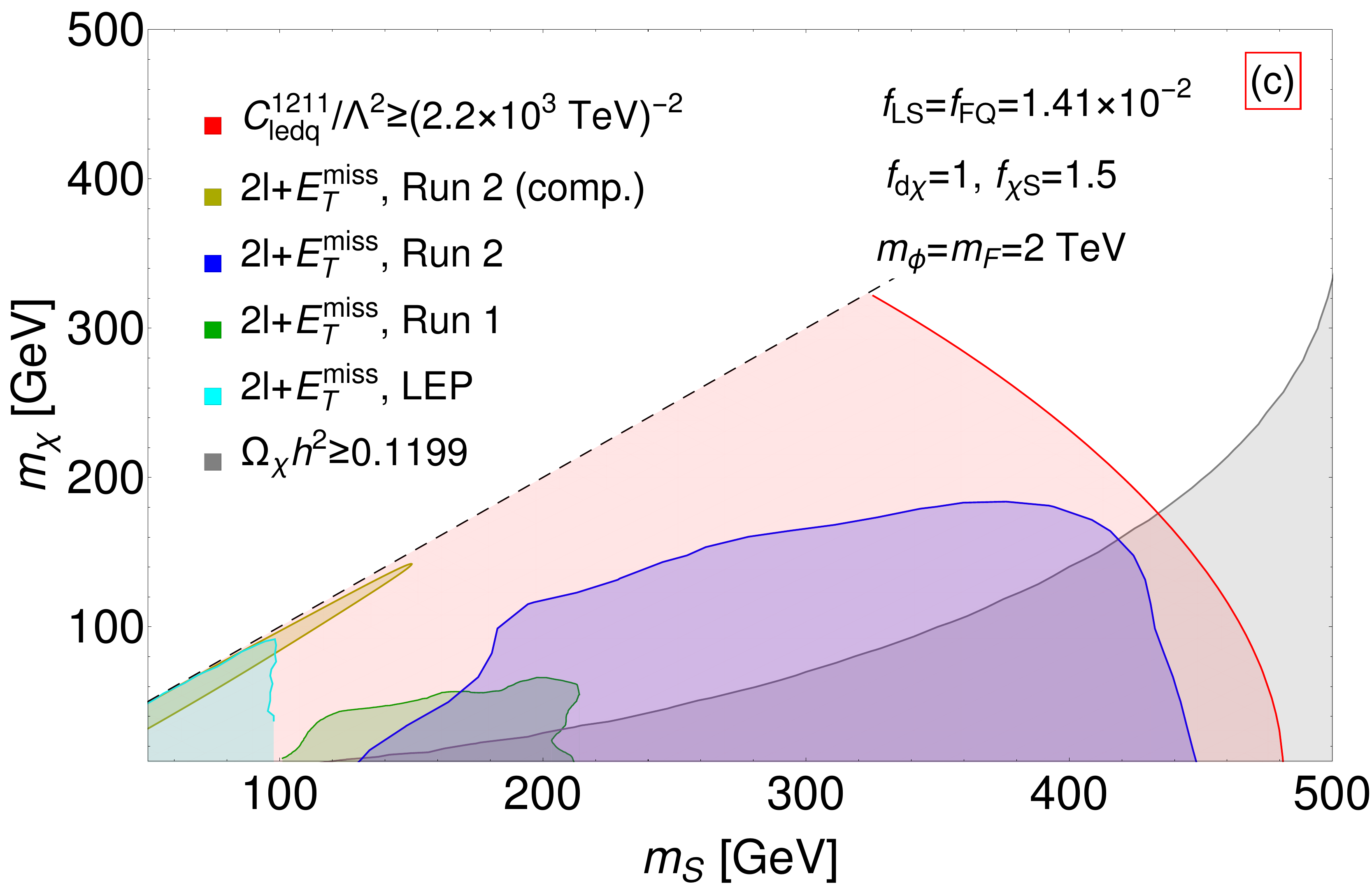}
\includegraphics[width=0.48\linewidth]{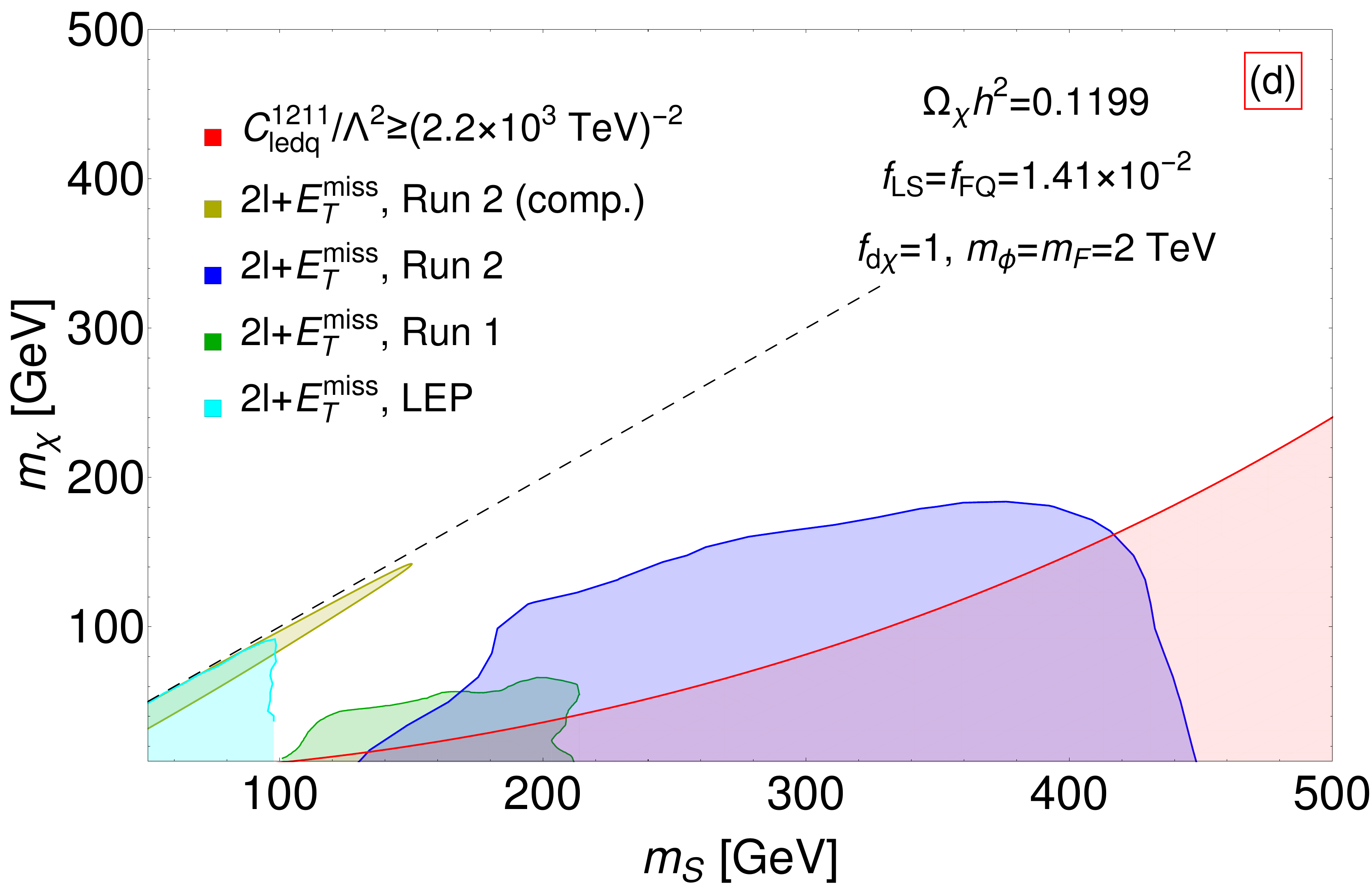}
     \caption{Combined sensitivities to the DM mass $m_\chi$ and the scalar mass $m_S$ for BM~(c) and BM~(d).
     The exclusions are obtained with the same experimental results as Fig.~\ref{fig:combined_flavor_conserving}, except for the current LFV constraint $C_{ledq}^{1211}/\Lambda^2< \left(2.2\times 10^3 \tev \right)^{-2}$~\cite{Fernandez-Martinez:2024bxg}.
     }
     \label{fig:combined_flavor_changing} 
\end{figure}

In Figs.~\ref{fig:combined_flavor_conserving} and \ref{fig:combined_flavor_changing}, we combine the sensitivities to the DM mass $m_\chi$ and scalar mass $m_S$ from DM relic density, direct detection, collider searches, and model-independent indirect constraints on the Wilson coefficients of the four-fermion operator $O_{ledq}^{\alpha\beta 11}$. 
For all benchmark scenarios in Table~\ref{tab:cases},  a large portion of parameter space with $m_\chi \lesssim 200\gev$ and $m_S \lesssim 450\gev$ is excluded by the existing searches at the LEP and LHC.

For the LFC operator, we consider the projected future sensitivity to Wilson coefficient, $C_{ledq}^{2211}/\Lambda^2< (12.3~\text{TeV})^{-2}$~\cite{Du:2020dwr, Du:2021rdg} from the neutrino NSI measurements, given that the currently allowed value is consistent with zero as mentioned in Sec.~\ref{sec:wilson}. For the LFV operator, we utilize the current constraint, $C_{ledq}^{1211}/\Lambda^2< \left(2.2\times 10^3 \tev \right)^{-2}$~\cite{Fernandez-Martinez:2024bxg} derived from the cLFV searches for $\mu \to e$ conversion in nuclei, since our aim is to show that the new physics scale can be notably alleviated within the {\it dark loop} paradigm.

In the left panel of Fig.~\ref{fig:combined_flavor_conserving}, we consider $f_{\chi S}=2$, and require the DM relic density $\Omega_{\chi} h^2 \leq 0.1199$ to avoid an overabundance of DM, shown as the gray shaded region. Given the values of $f_{d\chi}$ and $m_\phi$, DM direct detection in the LZ experiment~\cite{LZ:2022lsv} rules out a narrow range of $m_S$ and $m_\chi$, as depicted in purple slash shading area. The sensitivity of DM direct detection in this case is rather weak, which is suppressed by the relic density of $\chi$ for a relatively large $f_{\chi S}$.
The exclusion by the LFC constraint on $C_{ledq}^{2211}/\Lambda^2$ from the neutrino NSI measurements is shown in red color. It is evident that this model-independent constraint exhibits better sensitivity, probing parameter regions beyond the reach of collider searches, and DM direct detection, especially in scenarios where $m_\chi$ and $m_S$ are close. 

In the right panel of Fig.~\ref{fig:combined_flavor_conserving}, the coupling $f_{\chi S}$ is not independent but determined by the DM relic density $\Omega_{\chi} h^2 = 0.1199$.
The coupling $f_{\chi S}$ decreases as $m_S$ becomes smaller. The purple slash shading band for $15\gev < m_\chi < 100\gev$ is ruled out by DM direct detection in the LZ experiment.
The region excluded by the model-independent LFC constraint is depicted in red, with the boundary corresponding to $f_{\chi S} \simeq 2$. 
One can see that the Yukawa coupling $f_{\chi S}\sim \mathcal{O}(1)$ can yield the correct relic density for the electroweak scale $m_\chi$ and $m_S$~\cite{Liu:2021mhn}. 
In this scenario, the most effective constraints are provided by LHC searches and DM direct detection.
For larger values of $m_S$ and $m_\chi$, however, the neutrino NSI measurements are more significant.

In Fig.~\ref{fig:combined_flavor_changing}, the exclusions are derived with the same experimental results as Fig.~\ref{fig:combined_flavor_conserving}, except for the current model-independent constraint on the Wilson coefficient of the LFV four-fermion operator $C_{ledq}^{12 11}/\Lambda^2 < (2.2\times 10^3 \tev)^{-2}$. As highlighted in Sec.~\ref{sec:model}, the model-independent constraint on the Wilson coefficient of LFV operator is more stringent and can probe smaller new physics couplings,
compared to that on the Wilson coefficient of the LFC operator. For illustration, we consider $f_{d\chi} = 1$, for which the colored mediator mass $m_\phi \gtrsim 1.76\tev$ is allowed by the current LHC jet(s)$+\slashed E_T$ searches. Taking $m_\phi =2\tev$, the DM direct detection in the LZ experiment~\cite{LZ:2022lsv} puts no constraints as shown in Fig.~\ref{fig:dd}. 
In the case of BM~(c) with $f_{\chi S}=1.5$, the prevailing cLFV constraint rules out most of the parameter space where $m_\chi \lesssim 300\gev$ and $m_S \lesssim 450\gev$, exceeding the sensitivity of the LHC searches, especially for the region where $m_\chi$ and $m_S$ are close. On the other hand, for BM~(d) with the relic density $\Omega_\chi h^2=0.1199$ being fixed, the Yukawa coupling $f_{\chi S}\sim \mathcal{O}(1)$ can also yield the correct relic density for the electroweak scale $m_\chi$ and $m_S$. The LHC searches remain the most sensitive probes of the fermion portal DM model for the scalar mass $m_S \lesssim 420\gev$. As $m_S$ increases or DM mass $m_\chi \gtrsim 200\gev$, the cLFV searches exhibit better sensitivity than the LHC searches.

\section{Conclusion}
\label{sec:conclusion}

In this work, we have studied the UV realization of the four-fermion operator $O_{ledq}^{\alpha\beta 11}$ in both  lepton-flavor-conserving (LFC) and lepton-flavor-violating (LFV) scenarios incorporating the Majorana dark matter (DM). Due to the $Z_2$ symmetry that stabilizes the DM, the four-fermion operator $O_{ledq}^{\alpha\beta 11}$ is firstly generated at one-loop level via box diagram, which could effectively mitigate the lower bounds on the new physics scale.

We investigated the interplay between the model-independent constraints on the Wilson coefficients of the four-fermion operator $O_{ledq}^{\alpha\beta 11}$ from the measurements of neutrino NSI in the next-generation neutrino oscillation experiments, charged-lepton-flavor-violation (cLFV) searches in $\mu \to e$ conversion, as well as DM relic density, direct detection, and collider searches in the context of fermion portal DM model with two mediators.
We obtained the projected future bound on the LFC coefficient, $C_{ledq}^{2211}/\Lambda^2< (12.3~\text{TeV})^{-2}$ from neutrino NSI, and the LFV constraint $C_{ledq}^{1211}/\Lambda^2< \left(2.2\times 10^3~\text{TeV} \right)^{-2}$ from $\mu \to e$ conversion.
In order to illustrate the complementarities, we considered four benchmark scenarios as outlined in Table~\ref{tab:cases}. The colored mediator has a typical mass around $2~\text{TeV}$ and a coupling to quark $f_{d\chi}$ being 1-2; these values satisfy the constraints from both DM direct detection and jets+$\slashed{E}_T$ searches at the LHC. 

In the cases of BM~(a) and BM~(c), where the Yukawa coupling $f_{\chi S}$ is considered as an independent parameter, the model-independent constraints on the Wilson coefficients $C_{ledq}^{22 11}$ and $C_{ledq}^{12 11}$ provide a complementary investigation of the parameter space of the DM mass $m_\chi$ and scalar mass $m_S$, which extends the reaches of collider searches in the diplepton+$\slashed{E}_T$ channel and DM direct detection, especially for the region where $m_\chi$ and $m_S$ are close.

In the cases of BM~(b) and BM~(d), the coupling $f_{\chi S}$ is determined by the requirement of DM relic density, $\Omega_\chi h^2 = 0.1199$. For electroweak scale DM and scalar, the LHC searches demonstrate greater sensitivity than the indirect constraints on the four-fermion operator, while the latter are more significant for larger values of $m_\chi$ and $m_S$.


\begin{acknowledgments}
GL expresses gratitude to Shao-Long Chen for the enlightening discussion.
XZ would like to thank Hao-Lin Li, Jian Tang, Jiang-Hao Yu, and Zhao-Huan Yu for helpful discussions.
GL and XZ are supported by the National Natural Science Foundation of China under Grant No.~12347105 and No.~12505127, the Guangdong Basic and Applied Basic Research Foundation (2024A1515012668). 
The work of J.L. is supported by the National Science Foundation of China under Grant No. 12235001, No. 12475103 and State Key Laboratory of Nuclear Physics and Technology under Grant No. NPT2025ZX11.
The work of X.P.W. is supported by National Science Foundation of China under Grant No. 12375095, and the Fundamental Research Funds for the Central Universities.
\end{acknowledgments}

\appendix

\section{Lepton \tf{$g-2$}{g-2}}
\label{app:g-2}

\begin{figure}[!htb]
\centering
\includegraphics[width=0.35\linewidth]{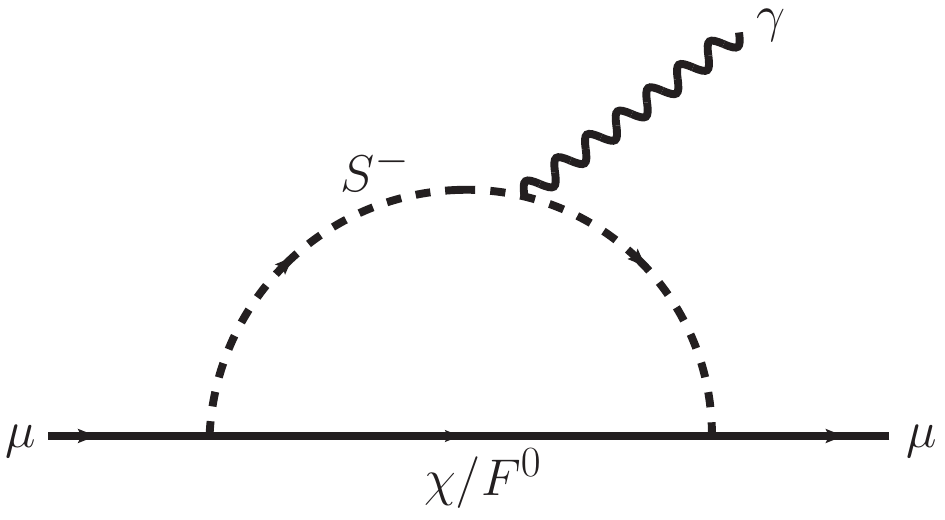} \hspace{0.05\linewidth}
\includegraphics[width=0.35\linewidth]{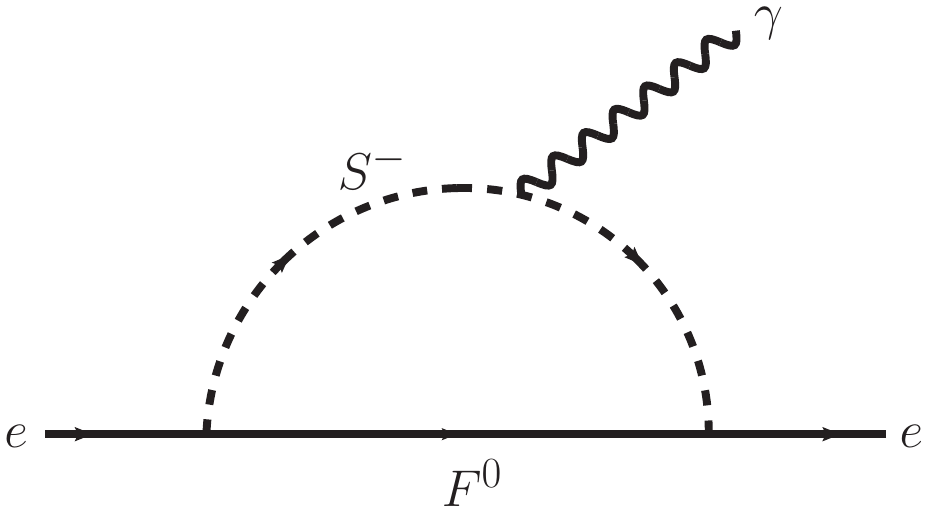}
\caption{Feynman diagrams for the muon and electron magnetic dipole moments generated at one-loop level.
}
\label{fig:muong2} 
\end{figure}

The lepton magnetic dipole moment $(g-2)$ can be induced by the couplings $f_{LS}$ and $f_{\chi S}$ at one-loop level. In Fig.~\ref{fig:muong2}, we depict the Feynman diagram with $F^0 $ or $\chi$ running in the loop.  

The corresponding contributions to muon $g-2$ is expressed as
\begin{align}
    \Delta a_{\mu}\equiv  a_{\mu}^{\text{exp}}-a_{\mu}^{\text{SM}}= \Delta a_{\mu}^F + \Delta a_{\mu}^{\chi}\;,
\end{align}
where $ \Delta a_{\mu}^F$ and $\Delta a_{\mu}^{\chi}$ denotes the contributions from the diagrams involving $F^0$ and $\chi$, respectively. For the electron $g-2$, we have
\begin{align}
    \Delta a_{e}\equiv  a_{e}^{\text{exp}}-a_{e}^{\text{SM}}= \Delta a_{e}^F \;,
\end{align}
which only includes the contribution from the diagram with $F^0$ given the lepton flavors $(\alpha,\beta)=(1, 2)$.

The piece $\Delta a_{\ell}^{F}$ $(\ell = e,\mu)$ is given by~\cite{Moroi:1995yh,Carena:1996qa} 
\begin{align}
\label{g2}
    \Delta a_{\ell}^{F}  &= -\frac{f_{L S}^2}{16 \pi^2} \frac{m_{\ell}^2}{m_S^2} g(x)\;,\quad g(x) \equiv \frac{1-6 x+3 x^2+2 x^3-6 x^2 \log x}{6(1-x)^4}\;,
\end{align}
where $x \equiv m_F^2 / m_S^2$, and $\Delta a_{\mu}^\chi$ can be obtained by taking $m_\ell = m_\mu$ and replacing $m_F \to m_\chi$ and $f_{LS} \to f_{\chi S}$. 
In the limits of $x\to 0$ and $x\to 1$, the loop function $g(x)$ approaches $1/6$ and $1/12$, respectively. Thus the contributions to the lepton $g-2$ are always negative~\cite{Liu:2021mhn}.

The experimental measurements of the muon $g-2$ yield~\cite{Muong-2:2025xyk} 
\begin{align}
   \Delta a_\mu=39(64)\times 10^{-11}\;,
\end{align}
by using the updated SM prediction~\cite{Aliberti:2025beg}, which can be either positive or negative.

There are still large uncertainties in the measurements of the electron $g-2$. The measurements using Cesium~\cite{Parker:2018vye} and Rubidium~\cite{Morel:2020dww} give
\begin{align}
   \Delta a_e(\text{Cs}) &= (-88 \pm 36) \times 10^{-14}, \nn\\
   \Delta {a}_e(\text{Rb}) &= (48 \pm 30) \times 10^{-14}\;.
\end{align}
Following Ref.~\cite{Liu:2021mhn}, we take the 95\% C.L. bounds on $\Delta a_e(\text{Cs})$ and $\Delta {a}_e(\text{Rb})$, so that both of them allow for negative values. If future measurements favor the positive $\Delta a_e(\text{Rb})$ result, our model could provide viable explanations for the anomalous magnetic moments of electron and muon.

\section{DM direct detection via photon exchange}
\label{app:anapole}

\begin{figure}[!htb]
	\centering
\includegraphics[width=0.4\linewidth]{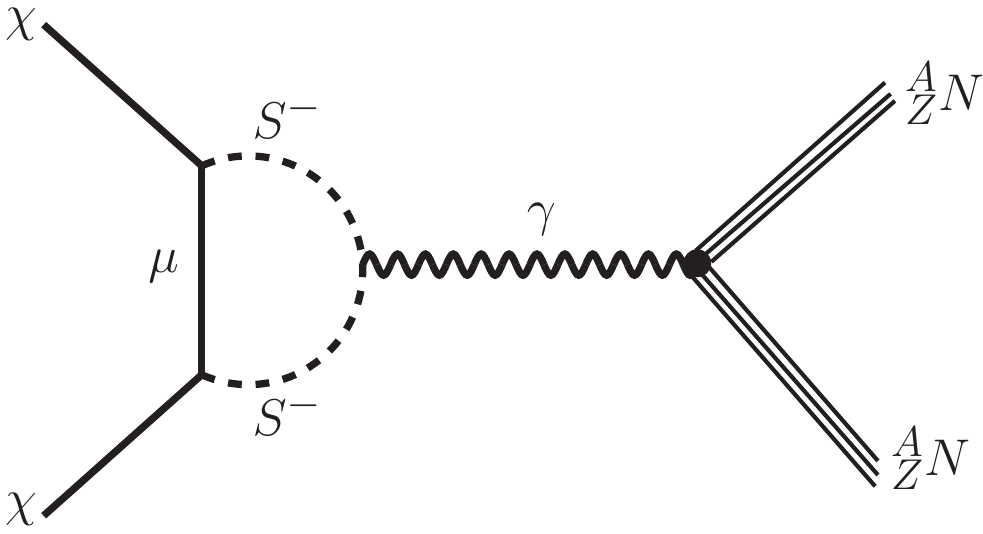}
	\caption{Representative Feynman diagram for DM-nucleus scattering at one-loop level via the exchange of photon.}
 \label{fig:exchangingphoton}	
\end{figure}

Besides the tree-level contribution to the DM direct detection discussed in Section.\ref{DD}, the DM-nucleus scattering can occur with the exchange of photon at the one-loop level~\cite{Bai:2014osa,Herrero-Garcia:2018koq}, as illustrated in Fig.~\ref{fig:exchangingphoton}. The interaction between the Majorana DM and photon can be described by the electromagnetic anapole momentum of the DM. Thus, the contribution to the SI DM-nucleon cross section is suppressed by the DM velocity square, which can be expressed by~\cite{Bai:2014osa}
\begin{align}
\label{formulaforDD}
    \sigma_{\rm SI}^{\rm{ana.}}=\frac{c_{\rm ana.}^2 e^2 Z^2}{2 \pi A^2} \frac{E_R^{\mathrm{ref}} m_p^2\left(m_T+m_\chi\right)^2}{m_T\left(m_p+m_\chi\right)^2}\;,
\end{align}
where $E_R^{\rm ref}$ denotes the reference value of the recoil energy $E_R \equiv q^2/(2m_T)$, and the loop factor is given by
\begin{align}
    c_{\rm ana.} = \frac{f_{\chi S}^2 e}{96 \pi^2 m_S^2}  \ln\left({m_{\mu}^2}/{m_S^2}\right)\;.
\end{align}
Here, $m_T$, $m_p$ and $m_\mu$ are the masses of the nucleus, proton and muon, respectively. $Z$ denotes the atomic number of the nucleus.  As an estimate, following Ref.~\cite{Bai:2014osa}, we take $E_R^{\rm ref} = 10~{\rm keV}$, and calculate $\sigma_{\rm SI}^{\rm{ana.}}=6.4\times 10^{-49}~{\rm cm}^2$ using liquid xenon as the target for $m_\chi = 50\gev$, $m_S = 100\gev$ and $f_{\chi S}=1$. Given the sensitivities of the DM direct detection experiments~\cite{LZ:2022lsv,PandaX:2022xas,XENON:2023cxc}, we obtain that it gives negligible contribution to the DM-nucleus cross section~\cite{Liu:2021mhn}.

\bibliographystyle{apsrev4-1}
\bibliography{reference}

\end{document}